\documentclass[sigconf]{acmart}

\usepackage{booktabs}
\usepackage{tabularx}
\usepackage{adjustbox}
\usepackage{pifont}
\usepackage{array,capt-of,rotating,multirow,hyperref}

\usepackage[most]{tcolorbox}
\usepackage{listings}
\usepackage{xcolor}
\definecolor{smCaseBorder}{RGB}{166,157,229}
\definecolor{smCaseTitle}{RGB}{213,207,255}
\newtcolorbox{smcase}[1]{
 enhanced,colback=white,colframe=smCaseBorder,
 colbacktitle=smCaseTitle,coltitle=black,
 fonttitle=\bfseries\footnotesize,title={#1},
 boxrule=0.85pt,arc=2mm,
 left=5pt,right=5pt,top=4pt,bottom=4pt,
 fontupper=\footnotesize,
 before skip=0pt,after skip=0pt
}

\setcopyright{none}
\acmYear{2027}
\copyrightyear{}
\acmDOI{}
\acmISBN{}
\acmConference[WWW '27]{the ACM Web Conference 2027}{May 10--14,
  2027}{Dublin, Ireland}
\begin{document}

\title{Beyond Profit Rankings: Identifying, Understanding, and Evaluating Smart Money in Prediction Markets}

%% Names, email, ORCID, and affiliations are kept in the source
%% and omitted from the review PDF. One \author command per person.
%% \email{name@institution.edu}
%% \orcid{0000-0000-0000-0000}

\author{Huaiyu Jia}
\affiliation{%
  \institution{The Hong Kong University of Science
  and Technology (Guangzhou)}
  \city{Guangzhou}
  \country{China}
}
\email{hjia351@connect.hkust-gz.edu.cn}

\author{Shuo Sun}
\affiliation{%
  \institution{The Hong Kong University of Science
  and Technology (Guangzhou)}
  \city{Guangzhou}
  \country{China}
}
\email{shuosun@hkust-gz.edu.cn}

\renewcommand{\shortauthors}{Author Name}

\begin{abstract}
Prediction markets have emerged as an important mechanism for aggregating information and trading beliefs about future events. The transparency of on-chain transactions has enabled investors to identify historically profitable wallets, commonly referred to as smart money, and replicate their trading activities through copy trading. However, despite the growing popularity of smart-money tracking, systematic research on these wallets' trading behavior, profitability persistence, and the economic feasibility of following their trades remains limited. To address this gap, we conduct a comprehensive empirical study of Polymarket, the most popular decentralized prediction market, by reconstructing wallet-level trading activities from on-chain transactions, position changes, market histories, and settlement outcomes. We first examine the distribution and persistence of historical profitability using statistical analysis and logistic regression, and further characterize heterogeneous trading strategies through multidimensional behavioral profiling and unsupervised clustering. Building on these findings, we develop an evidence-grounded multi-agent research framework that integrates wallet behavior, market conditions, and real-world information to generate interpretable and verifiable explanations of individual trades. Finally, we implement a real-time copy-trading system and evaluate its practical feasibility through order-book-aware paper trading and live on-chain execution. Our findings reveal substantial behavioral heterogeneity among historically profitable wallets and highlight the challenges of translating observed trading performance into replicable economic returns.
\end{abstract}

\keywords{Prediction Markets, Smart Money,
Copy Trading, On-Chain Analysis, Multi-Agent Systems}
%% Paste CCS XML and \ccsdesc lines from https://dl.acm.org/ccs .
%% CCS concepts are required for papers longer than two pages.
%% \begin{CCSXML}
%% <ccs2012>
%%  <concept>
%%   <concept_id>10002951.10003317.10003347.10003350</concept_id>
%%   <concept_desc>Information systems~Web searching and information discovery</concept_desc>
%%   <concept_significance>500</concept_significance>
%%  </concept>
%% </ccs2012>
%% \end{CCSXML}
%% \ccsdesc[500]{Information systems~Web searching and information discovery}

%% \keywords{keyword one, keyword two}

\maketitle

%% Tracks. Name the track, and the Web question it addresses, on
%% this first page. A Web dataset, API, or social network alone
%% is out of scope.
%%
%%   Brave New Ideas for the Future of the Web
%%   Web Economics and Digital Society
%%   Graph Algorithms and Modeling for the Web
%%   Responsible Web
%%   Search, Recommendation, and Retrieval-Augmented AI
%%   Security and Privacy
%%   Semantics and Knowledge
%%   Social Networks and Social Media
%%   Web Infrastructure and Agentic Systems
%%   User Modeling, Personalization and Agentic Web Users
%%   Web Mining, Multimedia and Multilingual Content Analysis
%%   Evaluation, Human Computation, and Resources
%%   Industry Track

\section{Introduction}

\begin{figure*}[htbp]
    \centering
    \includegraphics[width=0.95\textwidth]
    {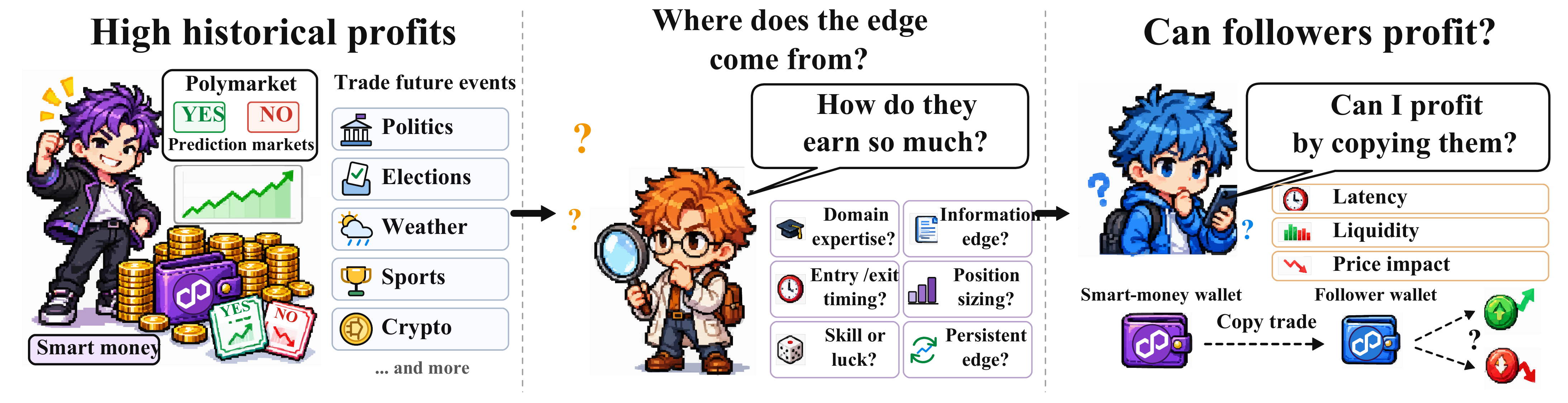}
    \caption{Motivation for studying smart money in
    prediction markets: concentrated historical profits,
    uncertain sources of trading advantages, and
    execution frictions faced by copy traders.}
    \Description{Three panels illustrating historical
    profits, potential sources of trading advantages,
    and copy-trading execution risks.}
    \label{fig:motivation}
\end{figure*}

In traditional financial markets, investors frequently seek to acquire informational advantages and generate excess returns (e.g., alpha) by identifying traders who exhibit high historical win rates and superior performance \cite{gruber1996another,zheng1999money}. In cryptocurrency markets, such participants, commonly referred to as \textbf{smart money} \cite{sapp2004momentum,akey2026polymarket}, are generally perceived to possess specialized knowledge, extensive trading experience, or informational advantages that enable them to achieve consistently superior investment performance. 
With the growth of blockchain-based applications,
including meme coins \cite{pumpfun,aveai},
tokenized equities \cite{xstocks,hyperliquid},
and prediction markets \cite{predictfun,polymarket},
transparent on-chain records enable investors to
track smart-money wallets and copy their
trades. As illustrated in Fig.~\ref{fig:motivation},
this raises three questions:
\textbf{(1)} Which wallets exhibit exceptional
historical profits?
\textbf{(2)} Do their returns reflect information
advantages, domain expertise, trading strategies,
or luck?
\textbf{(3)} Can followers replicate these returns
despite execution latency, limited liquidity,
and price impact?

\begin{figure}[t]
    \centering
    \includegraphics[width=\columnwidth]
    {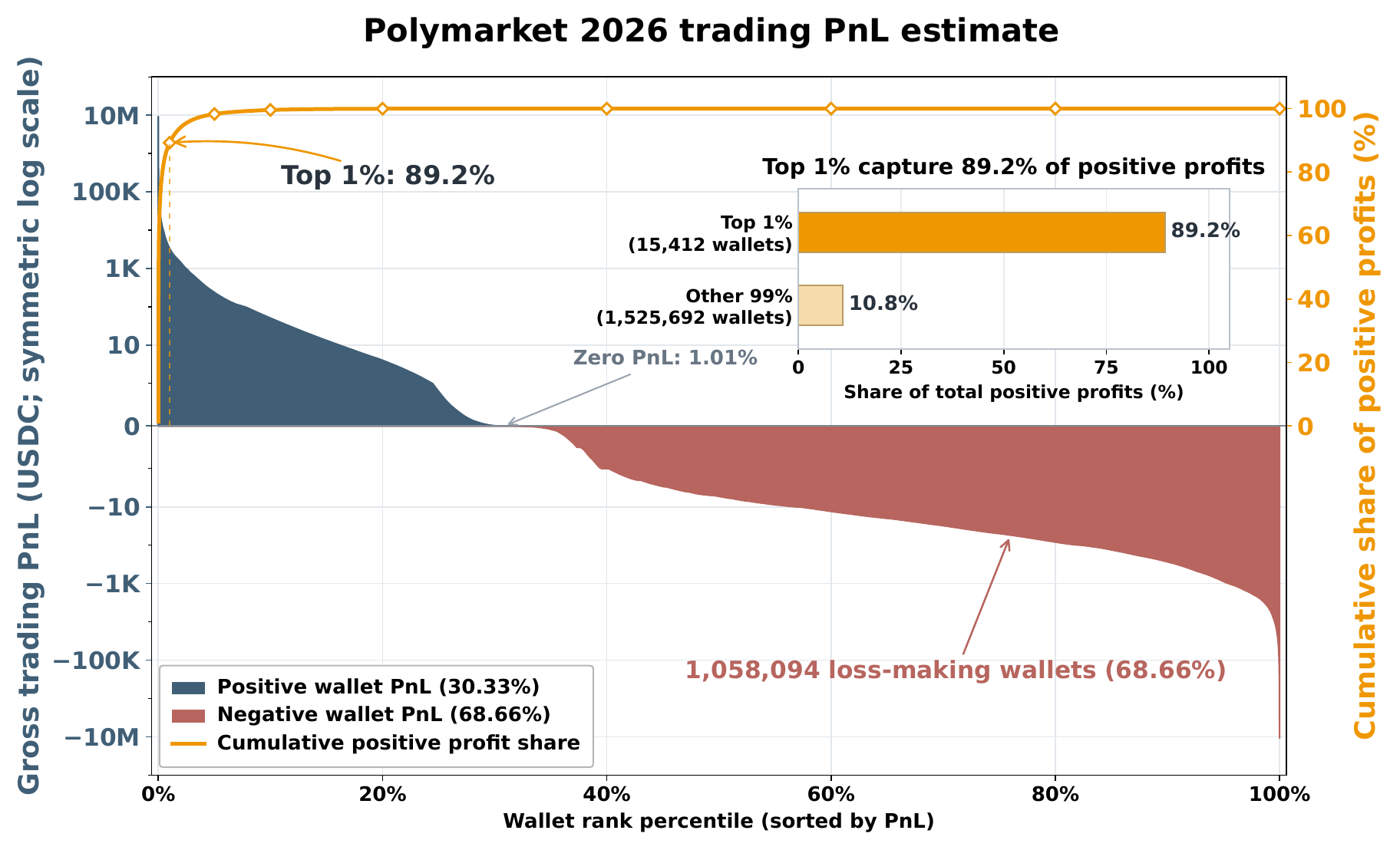}
    \caption{Distribution and concentration of wallet-level
    trading PnL on Polymarket. Wallets are ranked by estimated
    trading PnL in descending order. The blue and red regions
    represent positive and negative PnL, respectively, while
    the orange curve shows the cumulative share of positive
    profits. Among 1,541,104 wallets, 68.66\% exhibit
    negative PnL, whereas the top 1\% account for 89.2\%
    of total positive profits.}
    \label{fig:wallet_pnl_distribution}
\end{figure}

Decentralized prediction markets \cite{kalshi,predictfun}  offer a particularly suitable empirical setting for addressing these questions, as event-based settlement and traceable wallet-level transactions allow historical profitability to be examined in relation to both observable trading behavior and the execution constraints faced by followers. They allow participants to trade based on the outcomes of future events and dynamically adjust their positions before settlement. As a representative decentralized prediction market, Polymarket \cite{polymarket} employs an architecture combining off-chain centralized limit order book (CLOB) matching with Polygon's on-chain settlement, providing traceable data for wallet-level trading behavior research. In Figure \ref{fig:wallet_pnl_distribution}, 68.66\% of wallets were in a loss-making state, while the top 1\% of profitable wallets contributed 89.2\% of all positive returns. This significant concentration of profits prompted us to further investigate whether high-yield wallets possess a sustainable trading advantage and whether copy traders can replicate their profits.

Despite the growing popularity of smart-money tracking
and copy trading, several challenges remain.
First, existing platforms primarily identify smart-money
wallets through historical win rates, cumulative profits,
and trading activity \cite{hypertracker}, which cannot
reliably distinguish genuine trading skill from arbitrage
opportunities or random luck \cite{reichenbach2025exploring}.
Second, historical profitability alone provides limited
insight into traders' underlying strategies, including
market specialization, entry timing, and position management.
Although prediction markets offer explicit settlement rules
and verifiable outcomes \cite{umaoracle}, understanding
trading behavior requires jointly analyzing on-chain
transactions, market conditions, and real-world information.
Third, even profitable wallets may be difficult to follow,
as execution latency, limited liquidity, price impact,
and transaction fees can prevent copy traders from
achieving comparable returns.

To address these challenges, we conduct a systematic study
of smart money on Polymarket, combining on-chain data
analysis, statistical modeling, multi-agent research,
and real-world copy-trading experiments.
First, we reconstruct wallet-level trading activities
and position changes from Polymarket data and Polygon
transactions, and examine the distribution and persistence
of historical profitability through statistical analysis
and logistic regression.
Second, we characterize heterogeneous trading behavior
across six dimensions using rule-based profiling and
K-Means clustering, identifying distinct patterns in
trading activity, trade size, price preferences,
market specialization, position management, and
entry timing.
Third, we develop an evidence-grounded multi-agent
system that integrates wallet histories, market conditions,
and real-world evidence to generate traceable and
verifiable explanations of individual trades.
Finally, we implement an execution-constrained copy-trading
system and evaluate its practical feasibility through
order-book-aware paper trading and live on-chain execution,
examining the differences between source-wallet performance
and follower returns.

\section{Related Work}
%% Cite earlier papers of the authors in the third person.
%% Do not anonymize the citation or write ``omitted for review''.

\textbf{Smart Money and Copy Trading.} With the rapid development of meme coin and decentralized finance (DeFi) markets, smart-money tracking and copy trading have attracted increasing attention. Existing platforms typically identify potentially high-performing wallets based on historical win rates, cumulative profitability, and trading activity. For instance, Ave.ai \cite{aveai}, GMGN \cite{gmgn}, HyperTracker \cite{hypertracker}, Hyperliquid Analytics \cite{hyperliquidanalytics}, Polysights \cite{polysights}, and PolyBeats \cite{polybeats} provide services such as wallet profiling, profitability rankings, transaction monitoring, and trading signal analysis. Meanwhile, recent academic studies have begun to investigate on-chain trading behavior and its associated risks \cite{luo2026resisting,szwajcok2026meme,oliphant2026return}. However, the screening criteria and effectiveness of commercial smart-money services are not always publicly documented or independently validated, while existing academic research has largely focused on market manipulation and malicious trading activities. Systematic empirical investigations into the persistence of smart-money trading advantages, wallet-level trading behavior, and the practical replicability of copy-trading returns in prediction markets remain limited.

\noindent
\textbf{LLM-powered Agents in Blockchain.} In recent years, large language model (LLM)-powered agents have been increasingly adopted in blockchain and decentralized finance (DeFi) applications \cite{zhong2025detecting,muzammil2025poorest,niu2025natlm}. Prior studies have explored the use of LLM-based agents in cryptocurrency trading \cite{li2024cryptotrade,luo2025llm}, smart contract auditing \cite{wei2025advanced}, and the detection of cryptocurrency fraud and rug pulls \cite{wang2026cogagent}. Multi-agent systems support complex financial analysis through role division and collaborative reasoning. However, existing research primarily focuses on transaction decision-making, risk detection, and intent inference. Sufficient research remains needed on smart money analysis that combines on-chain behavior, historical returns, market conditions, and real-world information, while ensuring traceability and temporal consistency of evidence.

\begin{figure*}[t]
    \centering
    \includegraphics[
        width=0.98\textwidth,
        keepaspectratio
    ]{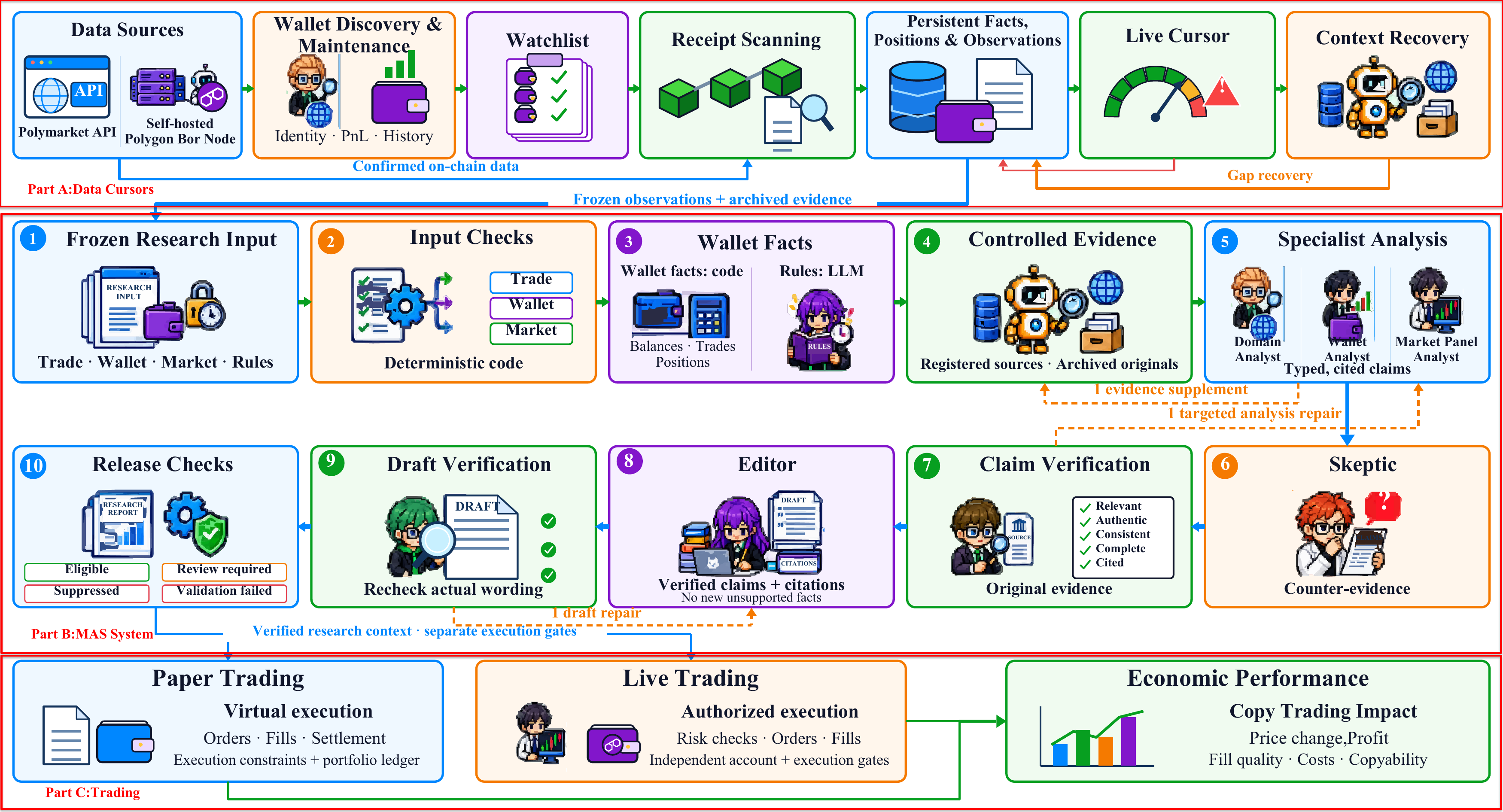}

    \caption{Overview of the proposed smart-money
    analysis framework. The system integrates
    Polymarket APIs and Polygon on-chain data
    for wallet discovery, continuous monitoring,
    and position reconstruction. An evidence-grounded
    multi-agent system analyzes wallet behavior,
    market conditions, and real-world evidence
    to generate verifiable trading explanations.
    Paper trading and live trading are handled
    through separate execution and risk controls.}

    \Description{System architecture showing continuous
    data ingestion, wallet monitoring, multi-agent
    trade research, evidence verification, and
    separate paper and live trading components.}

    \label{fig:system_overview}
\end{figure*}

\section{Smart-Money Identification and Analysis Framework}

In this section, we define the criteria for
identifying candidate smart-money wallets,
describe our continuous on-chain data collection
and position reconstruction pipeline, and introduce
an evidence-grounded multi-agent framework for
the smart money.

\subsection{Definition of Smart Money}
\label{sec:smart_money_definition}

In Polymarket, smart money generally refers to wallets perceived to exhibit persistent trading advantages in specific prediction-market domains. Because trading skill is not directly observable, we adopt an operational definition based on historical trading behavior and resolved event outcomes to identify candidate smart-money wallets. Specifically, for wallet $u$, market domain $s$, and evaluation cutoff time $T$, we examine the number of eligible events, directional win rate, net profit and loss (PnL), and profit factor over the preceding 180- and 90-day windows, together with the bidirectional-trading ratio and recent activity. Multiple trades associated with the same underlying event are consolidated into a single observation, and only events that have been formally resolved and verified before $T$ are included. The candidate smart-money indicator is defined as

\begin{equation}
\label{eq:sm_definition}
\begin{aligned}
\mathrm{SM}(u,s,T) = \mathbb{I}\Bigg[&
\bigwedge_{w\in\{180,90\}}
\bigl(N_w\ge n_w \land WR_w\ge r_w\\
&\qquad\land PNL_w>0\land PF_w\ge1.2\bigr)\\
&\qquad\land D_{180}\le0.10\land A_{90}\le30
\Bigg].
\end{aligned}
\end{equation}

Here, $\mathbb{I}[\cdot]$ denotes the indicator function, and $w$ denotes a lookback window measured in days. Table~\ref{tab:sm_criteria} summarizes the mathematical definitions and screening thresholds. $\mathcal{E}_w$ denotes the set of unique, verifiably resolved directional event clusters for wallet $u$ in domain $s$ during the lookback window; $\mathcal{R}_w$ denotes all verifiably resolved event clusters in that window; and $\mathcal{B}_{180}\subseteq\mathcal{R}_{180}$ denotes clusters involving trades on both sides. For event $e$, $d_e$ is the wallet's selected outcome, $y_e$ is the resolved winning outcome, $\pi_e$ is the attributable net PnL, and $\tau_e$ is the formal resolution time. The indicators implicitly depend on $(u,s,T)$. A zero denominator in $PF_w$ or incomplete evidence results in an undefined eligibility assessment rather than automatic qualification. These thresholds are prespecified screening criteria and do not themselves establish statistically significant predictive skill.

\begin{table}[t]
\centering
\caption{Operational Screening Criteria for Candidate
Smart-Money Wallets on Polymarket}
\label{tab:sm_criteria}

\footnotesize
\setlength{\tabcolsep}{2pt}
\renewcommand{\arraystretch}{1.4}

\begin{tabular*}{\columnwidth}
{@{\extracolsep{\fill}}llcc@{}}
\toprule
Metric & Mathematical Definition & 180d & 90d \\
\midrule

Eligible events
& $N_w=|\mathcal{E}_w|$
& $\ge 20$ & $\ge 10$ \\

Win rate
& $\displaystyle
WR_w=\frac{\sum_e\mathbb{I}(d_e=y_e)}{N_w}$
& $\ge 65\%$ & $\ge 60\%$ \\

Net PnL
& $\displaystyle PNL_w=\sum_e\pi_e$
& $>0$ & $>0$ \\

Profit factor
& $\displaystyle
PF_w=\frac{\sum_e\max(\pi_e,0)}
{\sum_e\max(-\pi_e,0)}$
& $\ge 1.2$ & $\ge 1.2$ \\

Bidirectional ratio
& $\displaystyle
D_{180}=\frac{|\mathcal{B}_{180}|}
{|\mathcal{R}_{180}|}$
& $\le 10\%$ & --- \\

Resolution age
& $A_{90}=T-\max_{e\in\mathcal{R}_{90}}\tau_e$
& --- & $\le 30\,\mathrm{d}$ \\

\bottomrule
\end{tabular*}

\vspace{2pt}
\parbox{\columnwidth}{
\scriptsize
\textit{Note:}
$w\in\{180,90\}$.
All sums $\sum_e$ are over
$e\in\mathcal{E}_w$.
$\mathcal{R}_w$ denotes resolved event clusters,
and $\mathcal{B}_{180}$ denotes
bidirectionally traded clusters.
Profit factor is undefined when its
denominator is zero.
}
\end{table}

\noindent\textbf{Statistical Validation.}
Following prior research distinguishing trading
skill from luck \cite{fama2010luck,barras2010false},
we assess the statistical reliability of our
identification framework using Poisson-binomial
tests, BH/BY multiple-testing corrections, and
dependence-aware null models
\cite{hong2013poisson,benjamini1995controlling,
benjamini2001control}.
Our analysis covers 1,101 candidate wallets
and 6,062,848 reconciled native orders.
We further conduct 2,000 Monte Carlo replications
per synthetic setting under varying event
probabilities, dependence structures, and
trading advantages.

Under correctly specified block-based null
models, BY correction yields false discovery
rates of 0.40--0.60\%, compared with
12.50--39.40\% when event dependence is
ignored. These results demonstrate the importance
of dependence-aware statistical inference and
support a two-stage identification framework
combining historical performance screening
with statistical skill assessment, rather than
classifying wallets solely by profitability
rankings or fixed numerical thresholds.

\subsection{Continuous Data Collection with Resumable Cursors}
We developed a continuous data collection pipeline that integrates the Polymarket API \cite{polymarketapi} with on-chain data. As shown in the \textbf{Part A of Fig \ref{fig:system_overview}}, we query the Polymarket API to retrieve category-specific profitability leaderboards, trading histories, and PnL snapshots (which are refreshed daily). Based on recent profitability and trading activity, we identify candidate wallets, verify their actual trading addresses, and perform deduplication, thereby dynamically maintaining a watchlist annotated with effective timestamps. Second, using a self-hosted Polygon Proof-of-Stake (PoS) full node running the Bor execution client \cite{polygonbor}, we continuously scan finalized blocks in near real time. By correlating trade execution events (OrderFilled), ERC-1155 token transfers \cite{erc1155}, and transaction receipts, we reconstruct wallet positions and identify position openings, additions, reductions, and exits. To ensure data collection continuity and recoverability, we employ persistent resumable data cursors that track API pagination progress, PnL snapshot timestamps, watchlist versions, and block heights. This mechanism enables recovery from interruptions, prevents duplicate processing of previously collected records, and explicitly tracks unresolved data gaps.

\begin{figure*}[htbp]
    \centering

    % (a) Market sector allocation
    \begin{minipage}[t]{0.32\textwidth}
        \centering
        \includegraphics[width=\linewidth]
        {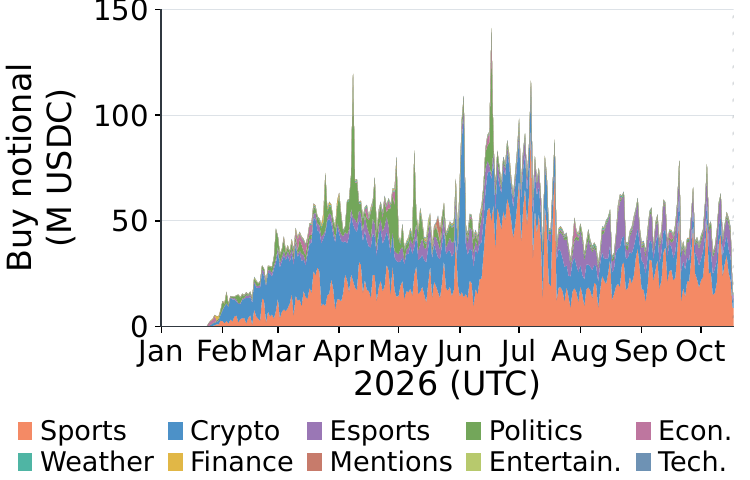}

        \par\smallskip
        {\footnotesize\textbf{(a) Market sector allocation}}
    \end{minipage}
    \hfill
    % (b) Buy-price distribution
    \begin{minipage}[t]{0.32\textwidth}
        \centering
        \includegraphics[width=\linewidth]
        {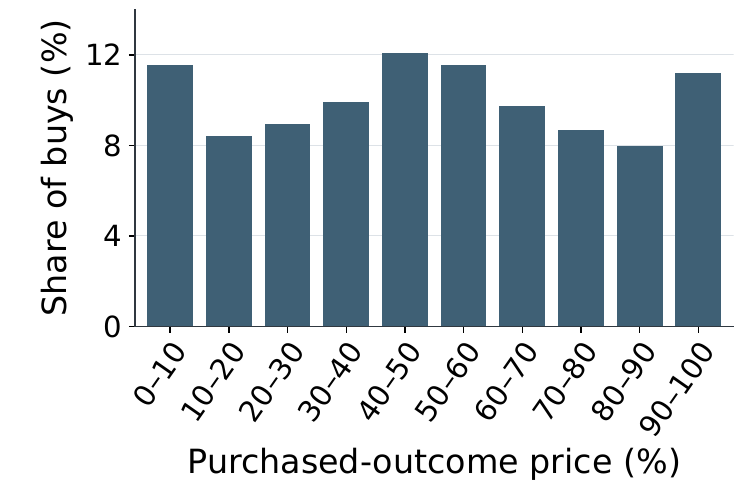}

        \par\smallskip
        {\footnotesize\textbf{(b) Buy-price distribution}}
    \end{minipage}
    \hfill
    % (c) Position management
    \begin{minipage}[t]{0.32\textwidth}
        \centering
        \includegraphics[width=\linewidth]
        {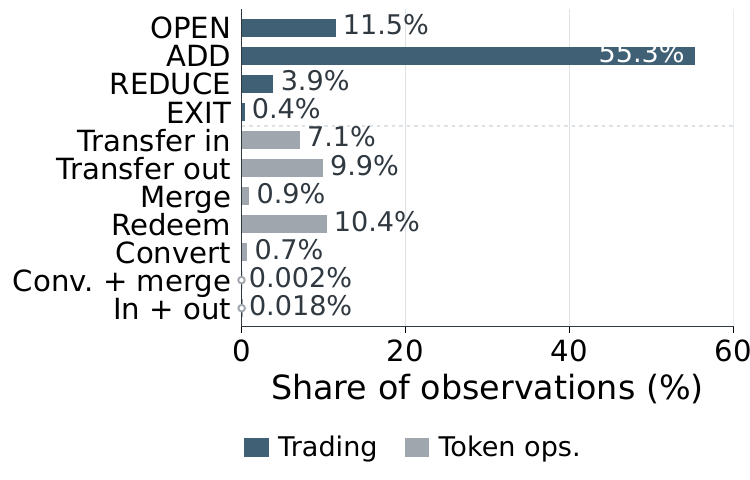}

        \par\smallskip
        {\footnotesize\textbf{(c) Position management}}
    \end{minipage}

    \caption{Descriptive characterization of trading
    behavior on Polymarket.
    (a) Daily buy notional across market sectors
    for the retrospectively selected top 1\%
    high-PnL wallets.
    (b) Distribution of purchased-outcome prices
    among captured candidate-wallet buy actions.
    (c) Distribution of captured position-management
    actions and non-trading token operations.
    The panels use different samples and denominators
    and are not directly comparable.}

    \label{fig:smart_money_behavior}
\end{figure*}

\subsection{Evidence-Grounded Multi-Agent Smart-Money Analysis}
\label{sec:mas_analysis}

To investigate the trading behavior of smart-money
wallets and their potential informational advantages,
we develop an evidence-grounded multi-agent research
framework, as illustrated in \textbf{Part B of Fig \ref{fig:system_overview}},
\textbf{Steps~\ding{172}--\ding{174}} freeze trade,
wallet, and market information, reconstruct transactions
and position changes through deterministic procedures,
and employ a Rules Analyst to interpret market resolution
conditions. \textbf{Step~\ding{175}} retrieves and
archives external evidence from controlled sources,
preserving original content, provenance, and timestamps
for subsequent analysis.

\par\noindent
\textbf{Step~\ding{176}: Specialist Multi-Agent Analysis.}
The framework employs specialized agents sharing
the same underlying LLM but operating with role-specific
prompts. \textbf{(1) Wallet Analyst} examines historical
transactions, account profitability, and position records
to characterize trading frequency, entry size, buy/sell
preferences, and position management. It also identifies
deviations from historical behavior, such as unusually
large purchases or repeated accumulation.
\textbf{(2) Domain Analyst} selects domain-specific
agents for politics, sports, esports, and other market
categories to investigate real-world events that may
affect market outcomes, distinguishing verified facts
from conditional interpretations and unverified
hypotheses. \textbf{(3) Market Panel Analyst} examines
price movements, bid-ask spreads, and order-book
liquidity to assess market conditions and execution
constraints. Each agent produces structured claims
with evidence references and temporal scope rather
than direct trading recommendations.

\par\noindent
\textbf{Steps~\ding{177}--\ding{178}} introduce a
\textbf{Skeptic Agent} and a \textbf{Claim} Verifier to critically
examine these analyses. The Skeptic challenges
unsupported interpretations of historical profitability,
trading motives, and informational advantages, while
the Claim Verifier checks claims against original
evidence, citation dependencies, and temporal
consistency. In \textbf{Steps~\ding{179}--\ding{181}},
an \textbf{Editor} organizes verified claims into a research
report, followed by draft verification and release
checks. This process integrates wallet behavior,
market conditions, and real-world evidence into
traceable trading explanations while keeping research
conclusions separate from actual copy-trading
authorization.

\subsection{Descriptive Analysis}
\label{sec:descriptive_analysis}

To characterize trading behavior on Polymarket, we conduct a descriptive analysis along three dimensions: market sector allocation, buy-price distribution, and position management. For sector allocation, we analyze the daily buy notional of 15,412 wallets retrospectively ranked in the top 1\% by estimated PnL among 1,541,104 evaluable wallets in 2026 through October 17; for buy-price distribution and position management, we use the October 6--13, 2026 observation window from 1,681 identified candidate wallets, covering 971 trading wallets with 3,380,953 buy/sell actions and 985 observed wallets with 4,763,410 total observations, respectively.

% ============================================================
% RQ1: Market Sector Allocation
% ============================================================

\subsubsection{RQ1: How Is Trading Activity Distributed Across Prediction Market Sectors?}
\label{sec:rq1_sector_allocation}

We first examine the market participation of 15,412 wallets retrospectively ranked in the top 1\% by historical PnL. As shown in Fig.~\ref{fig:smart_money_behavior}(a), sports and cryptocurrency markets account for substantial shares of daily buy notional, while esports and politics contribute noticeably during certain periods. Trading activity increases sharply during June and July, particularly in sports markets, and sectoral contributions vary considerably over time. Importantly, buy notional includes repeated purchases and position additions rather than net capital investment. Moreover, retrospective wallet selection does not establish ex-ante domain-specific expertise.

\noindent\textbf{Finding RQ1.} Historically high-PnL wallets exhibit uneven and time-varying trading activity across market sectors, with sports and cryptocurrency markets accounting for substantial buy notional. These patterns characterize market participation rather than verified sector-specific trading skill.

% ============================================================
% RQ2: Buy-Price Distribution
% ============================================================

\subsubsection{RQ2: At What Prices Do Candidate Smart-Money Wallets Trade?}
\label{sec:rq2_buy_prices}

We next examine the execution prices of purchased outcome shares among candidate smart-money wallets. Fig.~\ref{fig:smart_money_behavior}(b) shows that buy actions are distributed across the full price range rather than concentrated in a single interval. Specifically, purchases priced between 40\% and 60\% account for 23.63\% of all captured buys, while 11.55\% occur below 10\% and 11.19\% occur between 90\% and 100\%. These prices refer to the purchased YES or NO outcomes, rather than exclusively to YES probabilities. Furthermore, the distribution is based on transaction counts rather than buy notional and therefore does not directly measure capital allocation or risk preferences.

\noindent\textbf{Finding RQ2.} Smart-money wallets purchase outcome shares across diverse price intervals, including both low and high priced outcomes, the observed buying behavior does not exhibit an exclusive preference for either unlikely or highly probable outcomes.

% ============================================================
% RQ3: Position Management
% ============================================================

\subsubsection{RQ3: How Do Candidate Smart-Money Wallets Manage Their Positions?}
\label{sec:rq3_position_management}

Finally, we characterize wallet position management by combining on-chain trade executions, ERC-1155 transfers, and reconstructed position states. As shown in Fig.~\ref{fig:smart_money_behavior}(c), position additions (ADD) dominate captured observations at 55.3\%, followed by openings (OPEN) at 11.5\%, while reductions (REDUCE) and exits (EXIT) account for only 3.9\% and 0.4\%, respectively. Non-trading operations are also substantial: transfers in, transfers out, and redemptions account for 7.1\%, 9.9\%, and 10.4\% of observations. Moreover, the limited observation window prevents interpreting the low EXIT proportion as evidence of holding positions until resolution.

\noindent\textbf{Finding RQ3.} Candidate smart-money wallets exhibit a tendency toward incremental position accumulation rather than one-off entries. Position additions account for 55.3\% of captured observations, nearly five times the share of initial openings (11.5\%), suggesting that these wallets may favor scaling
into existing positions through repeated purchases, which indicates a potential preference for gradual position building.

\begin{table*}[htbp]
\centering
\caption{Six-dimensional behavioral profiles with
representative wallets and transactions.}
\label{tab:smart_money_behavior_profiles}

\begingroup
\fontsize{7.5}{8.8}\selectfont
\setlength{\tabcolsep}{1.15pt}
\renewcommand{\arraystretch}{1.07}

\begin{tabular*}{\textwidth}{
@{\extracolsep{\fill}}l l l l l c c c@{}}

\toprule
\textbf{Dim.} &
\textbf{Rule-based labels} &
\textbf{K-Means group ($n$; median)} &
\textbf{Wallet} &
\textbf{Tx} &
\textbf{Sil.} &
\textbf{Event ARI} &
\textbf{Time ARI} \\
\midrule

% ==============================
% A: ACTIVITY
% ==============================

\multirow{3}{*}{
\shortstack[l]{
\textbf{A. Activity}\\
$N=602$, $K=3$}} &
\multirow{3}{*}{
\shortstack[l]{
L/M/H: 200/201/201\\
33.2/33.4/33.4\%}} &
\mbox{\textbf{A-0} Multi-day; 364; 17.1/d} &
\href{https://polygonscan.com/address/0x5415c298dde8c84f4ba726d36b411df6e9a72a6e}{0x5415..2a6e} &
\href{https://polygonscan.com/tx/0xcffe62269d85b009578f5617b57b82befdb3a6c7a38fccbfb3d9818ace51bdbf}{0xcffe..bdbf} &
\multirow{3}{*}{0.496} &
\multirow{3}{*}{0.718} &
\multirow{3}{*}{0.513} \\

& &
\mbox{\textbf{A-1} Low; 131; 2.29/d} &
\href{https://polygonscan.com/address/0xe885f33c65fc6c244a297e9df9361b35975995b7}{0xe885..95b7} &
\href{https://polygonscan.com/tx/0x592f4ed7d368a90b0f43c0504f3957110adce7990cbe4ad5e13cad9dcb69008e}{0x592f..008e} &
& & \\

& &
\mbox{\textbf{A-2} Ultra-active; 107; 1,076/d} &
\href{https://polygonscan.com/address/0xc87836bef73740dd2206c05628e9572f59faa529}{0xc878..a529} &
\href{https://polygonscan.com/tx/0xf5f34271dd461b81ea4de02eaffa85ec0c3682bf501086bf334efcf450f69279}{0xf5f3..9279} &
& & \\

\midrule

% ==============================
% B: TRADE SCALE
% ==============================

\multirow{2}{*}{
\shortstack[l]{
\textbf{B. Scale}\\
$N=548$, $K=2$}} &
\multirow{2}{*}{
\shortstack[l]{
Large/turn.: 183/196\\
33.4\% (548/587)}} &
\mbox{\textbf{B-0} Large; 259; \$582/entry} &
\href{https://polygonscan.com/address/0x401ee31e9ebf9ab9f6315cd95faca5f950436fc9}{0x401e..6fc9} &
\href{https://polygonscan.com/tx/0x22e168ffbd88c68ee163959a203e5ad8e5ece2dbbe2c7dae1e454b7e2f68c8f1}{0x22e1..c8f1} &
\multirow{2}{*}{0.327} &
\multirow{2}{*}{0.814} &
\multirow{2}{*}{0.672} \\

& &
\mbox{\textbf{B-1} Small; 289; \$35.84/entry} &
\href{https://polygonscan.com/address/0x777fae71d2ff9ec48a1213d48ba1d9d91024a1bb}{0x777f..a1bb} &
\href{https://polygonscan.com/tx/0x07af9203e7ea5ed67264ab7c6f984f2e572ddd14e141adde16c8c7cda41a8ef9}{0x07af..8ef9} &
& & \\

\midrule

% ==============================
% C: PURCHASE PRICE
% ==============================

\multirow{2}{*}{
\shortstack[l]{
\textbf{C. Price}\\
$N=587$, $K=2$}} &
\multirow{2}{*}{
\shortstack[l]{
L/H/Mix: 29/235/166\\
4.9/40.0/28.3\%}} &
\mbox{\textbf{C-0} Mid-price; 335; 57.5\% [0.3,0.7)} &
\href{https://polygonscan.com/address/0x9a4cbaf3ba914b38ca92cc968cb6104c002998d0}{0x9a4c..98d0} &
\href{https://polygonscan.com/tx/0xc7dadb8e56aa00f60d46781b6863a5672b392585147299a47fd4139fb9d86582}{0xc7da..6582} &
\multirow{2}{*}{0.361} &
\multirow{2}{*}{0.751} &
\multirow{2}{*}{0.511} \\

& &
\mbox{\textbf{C-1} Very-high; 252; 45.5\% $\geq$0.9} &
\href{https://polygonscan.com/address/0x48185887c8dc95de60ee89722f1d0ee7894cbf0b}{0x4818..bf0b} &
\href{https://polygonscan.com/tx/0x09582bc3249736688f5b2e55ba67805dae2a2a070204ffe1f70c6a347bbfabc9}{0x0958..abc9} &
& & \\

\midrule

% ==============================
% D: MARKET FOCUS
% ==============================

\multirow{3}{*}{
\shortstack[l]{
\textbf{D. Focus}\\
$N=557$, $K=3$}} &
\multirow{3}{*}{
\shortstack[l]{
Focus/div.: 286/328\\
51.3/58.9\%}} &
\mbox{\textbf{D-0} Sector/multi-evt.; 204; 48 evts} &
\href{https://polygonscan.com/address/0x37461eff876736bf0363647db4b9c319c48bc193}{0x3746..c193} &
\href{https://polygonscan.com/tx/0x0e3206af0f76179b5c1097d54e33870271e488e59b5b6657b55b78925f3f0e99}{0x0e32..0e99} &
\multirow{3}{*}{0.413} &
\multirow{3}{*}{0.631} &
\multirow{3}{*}{0.472} \\

& &
\mbox{\textbf{D-1} Event-focused; 154; HHI 0.359} &
\href{https://polygonscan.com/address/0x392d557de378926143df5f72d8ddaef3f0d221a1}{0x392d..21a1} &
\href{https://polygonscan.com/tx/0xe5af8cc87f7c686981f41829eb1da8ecf3830eea359715c76cc4e64a854cd2a1}{0xe5af..d2a1} &
& & \\

& &
\mbox{\textbf{D-2} Broad/multi-evt.; 199; 8 sect.} &
\href{https://polygonscan.com/address/0x1ede425ac39b1286e42174e8f97dfa0ea76b46e8}{0x1ede..46e8} &
\href{https://polygonscan.com/tx/0x23756bebaa99b2641f77b995efad82ed1d9ffeeab20910782c53df7c6f92a565}{0x2375..a565} &
& & \\

\midrule

% ==============================
% E: POSITION MANAGEMENT
% ==============================

\multirow{2}{*}{
\shortstack[l]{
\textbf{E. Position}\\
$N=479$, $K=2$}} &
\multirow{2}{*}{
\shortstack[l]{
S/R/E: 124/40/184\\
25.9/8.4/48.8\%}} &
\mbox{\textbf{E-0} Fewer changes; 335; 0\% red.} &
\href{https://polygonscan.com/address/0xb6fbce093cdd139858c44148a6598d8ec028c038}{0xb6fb..c038} &
\href{https://polygonscan.com/tx/0xe11d71b13afb7702c36b4982d5c582cc007965069cd21e3055e14b9f7e48e5de}{0xe11d..e5de} &
\multirow{2}{*}{0.426} &
\multirow{2}{*}{0.747} &
\multirow{2}{*}{0.522} \\

& &
\mbox{\textbf{E-1} More changes; 144; 26.4\% red.} &
\href{https://polygonscan.com/address/0xdadbf9e1df1b8d7a184a0d6ab9c83b2337b61870}{0xdadb..1870} &
\href{https://polygonscan.com/tx/0x7c003bd64c4664ba31722471921fc87c598402a952c66a836f0f1fd29d7e91ea}{0x7c00..91ea} &
& & \\

\midrule

% ==============================
% F: ENTRY TIMING
% ==============================

\multirow{2}{*}{
\shortstack[l]{
\textbf{F. Timing}\\
$N=418$, $K=2$}} &
\multirow{2}{*}{
\shortstack[l]{
E/L/M: 44/182/176\\
10.5/43.5/42.1\%}} &
\mbox{\textbf{F-0} Late-tilted; 180; 88.9\% late} &
\href{https://polygonscan.com/address/0x686880eea810fa9141be46a1acec9eee41755198}{0x6868..5198} &
\href{https://polygonscan.com/tx/0x9eb6be8d5d900764a8ef4c4662103ae0c41948af064f594e81c6496baee5d9fb}{0x9eb6..d9fb} &
\multirow{2}{*}{0.501} &
\multirow{2}{*}{0.796} &
\multirow{2}{*}{0.687} \\

& &
\mbox{\textbf{F-1} Earlier-tilted; 238; 36.5\% early} &
\href{https://polygonscan.com/address/0x9690d9f51b618222c29567c3b39aac2935a5f2d5}{0x9690..f2d5} &
\href{https://polygonscan.com/tx/0xd4cb5740123eddb08107b20d4c8054d2710bcfd1d0d6f4b9fbf5a1608296c49d}{0xd4cb..c49d} &
& & \\

\bottomrule
\end{tabular*}

\vspace{3pt}
\parbox{\textwidth}{
\fontsize{7}{8.4}\selectfont
\emph{Notes.}
A fixed cohort contains 1,101 candidate wallets;
$N$ is the number evaluable per dimension and
$K$ is the selected cluster count.
Rule-based labels may overlap; their proportions
have label-specific denominators
(high turnover: 587; verified exits: 377;
otherwise the dimension's $N$).
Sil. is silhouette;
Event ARI is parent-event bootstrap stability;
Time ARI is cross-week clustering agreement,
not profit prediction.
Each wallet and transaction hash links to PolygonScan.
}
\endgroup
\end{table*}

\section{Experiment}
In this section, we evaluate the persistence of
historical wallet profitability, characterize
smart-money trading behavior through clustering,
examine the interpretability of our multi-agent
framework, and assess the execution feasibility
and economic performance of paper and live
copy trading.

\subsection{Experimental Setup} The multi-agent system uses a Qwen3.8-27B model \cite{qwen38} served by SGLang on two NVIDIA RTX 5090 GPUs (TP=2). Agents share model weights but use role-specific prompts, with temperature set to 0.2, Thinking Mode disabled, and a 32,768-token context. The system is implemented in Python 3.12 using our MasEngine orchestrator. Separately, wallet behavior clustering is performed on a CPU server equipped with two AMD EPYC 9554 processors. DuckDB 1.5.5 processes verified trading and market data, while scikit-learn 1.9.1 trains six independent K-Means models. We analyze 1,101 candidate wallets using October 6–13, 2026 for training and October 13–20 for validation (UTC). Detailed experimental configurations and hyperparameters are provided in the Appendix \ref{experimental-parameters}.

\subsection{Profitability Persistence Analysis}

\textbf{Short-Term Predictive Power of Historical Profits.}
\label{sec:profit_persistence}
To examine the predictive power of historical profits for future performance, we constructed a weekly profit panel based on 1,101 candidate wallets, of which 1,088 wallets formed 3,207 complete weekly pairings. Let $\pi_{i,t}$ represent the PnL of wallet $i$ in week $t$. The profitability of the next week is defined as:
\begin{equation}
\label{eq:weekly_target}
Y_{i,t+1}=\mathbb{I}(\pi_{i,t+1}>0).
\end{equation}

We use binary logistic regression to predict the probability of profitability for the next week:
\begin{equation}
\label{eq:weekly_probability}
\begin{aligned}
\hat p_{i,t+1}
&=\Pr(Y_{i,t+1}=1\mid X_{i,t})\\
&=\sigma(\eta_{i,t})
=\frac{1}{1+\exp(-\eta_{i,t})}.
\end{aligned}
\end{equation}

To evaluate the incremental predictive value of historical information, we construct three models: M0 uses the overall profitability during the training period, M1 incorporates the profitability of the previous week, M2 further incorporates the number of transactions and the size of profit and loss:
\begin{equation}
\label{eq:weekly_models}
\begin{aligned}
\mathrm{M0}:\quad
&\hat p_{i,t+1}=\bar Y_{\mathrm{train}},\\
\mathrm{M1}:\quad
&\eta_{i,t}=\beta_0^{(1)}
+\beta_1^{(1)}D_{i,t},\\
\mathrm{M2}:\quad
&\eta_{i,t}=\beta_0^{(2)}
+\beta_1^{(2)}D_{i,t}\\
&\qquad+\beta_2^{(2)}Z^N_{i,t}
+\beta_3^{(2)}Z^P_{i,t}.
\end{aligned}
\end{equation}

The mathematical definitions of each variable are shown in Table~\ref{tab:logit_variables}. Continuous variables are standardized using the mean and standard deviation of the training set, to avoid data leakage. The model uses maximum likelihood estimation and time-leave validation, ensuring that training data is available before testing. We use the Brier Score to measure probability prediction error; a lower score indicates better prediction performance.

\begin{table}[t]
\centering
\caption{Definitions of Logistic Regression Variables}
\label{tab:logit_variables}
\footnotesize
\setlength{\tabcolsep}{3pt}
\renewcommand{\arraystretch}{1.2}
\begin{tabular}{@{}lp{0.76\columnwidth}@{}}
\toprule
Variable & Definition\\
\midrule
$\pi_{i,t}$ & $P_i(t_{\mathrm{end}})-P_i(t_{\mathrm{start}})$; weekly position PnL change \\
$D_{i,t}$ & $\mathbb{I}(\pi_{i,t}>0)$; previous-week profitability \\
$N_{i,t}$ & Increment in the previous week's reported trade count \\
$Z^N_{i,t}$ & $\displaystyle \frac{\ln(1+N_{i,t})-\mu_N}{s_N}$; standardized trading activity \\
$Z^P_{i,t}$ & $\displaystyle \frac{\ln(1+|\pi_{i,t}|)-\mu_P}{s_P}$; standardized absolute PnL \\
\bottomrule
\end{tabular}
\vspace{2pt}
\parbox{\columnwidth}{\scriptsize \textit{Note:} $P_i$ denotes provider-reported cumulative position PnL. $\mu$ and $s$ are computed from training data only. Weekly PnL includes realized and unrealized Changes.}
\end{table}

\begin{table}[t]
\centering
\caption{Out-of-Sample Weekly Profitability Prediction}
\label{tab:weekly_logit}
\footnotesize
\setlength{\tabcolsep}{3pt}
\renewcommand{\arraystretch}{1.15}
\begin{tabular*}{\columnwidth}{@{\extracolsep{\fill}}llrr@{}}
\toprule
Model & Features & \shortstack{Full\\Sample} & \shortstack{Excl.\\Zero-PnL} \\
\midrule
M0 & Base rate & 0.2493 & 0.2480 \\
M1 & Previous profit & 0.2306 & \textbf{0.2459} \\
M2 & Profit + Activity + PnL & \textbf{0.2278} & 0.2470 \\
\midrule
M2 AUC & --- & 0.681 & 0.577 \\
\multicolumn{2}{l}{Test observations} & 2,156 & 1,625 \\
\bottomrule
\end{tabular*}
\vspace{2pt}
\parbox{\columnwidth}{\scriptsize \textit{Note:} Brier scores (lower is better). }
\end{table}

As shown in Table~\ref{tab:weekly_logit}, the Brier score of M2 in the full sample decreased from 0.2493 to 0.2278, representing an improvement of approximately 8.6\% relative to the benchmark, with an AUC of 0.681, indicating that historical profitability and trading activity have short-term predictive value. Excluding observations of zero returns from the previous week, M1 achieved the lowest Brier Score (0.2459), while M2 only reached 0.2470, with the AUC decreasing to 0.577, indicating that adding trading features does not always improve predictive performance.

\noindent\textbf{Finding.}
Candidate smart-money wallets exhibit short-term profitability
persistence, with previously profitable wallets showing
2.63 times higher odds of remaining profitable in the following week.
However, this advantage weakens after excluding zero-PnL
observations, suggesting limited robustness.

% Requires: booktabs, multirow, hyperref

\subsection{Smart-Money Trading Behavior Clustering}

\label{subsec:sm_style_clustering}

To characterize the heterogeneity of trading behavior among Polymarket candidate smart wallets, we conducted a six-dimensional behavioral analysis of 1,101 wallets in a fixed queue, covering trading activity, trading scale, entry price preference, market focus, position management, and entry timing. Behavioral features for each dimension were constructed based on deduplicated on-chain orders, historical market information, and verifiable changes in holdings. The experiment used October 6–13, 2026 as the training window and October 13–20, 2026 as an independent validation window. Due to the different requirements for transaction records and historical evidence for each dimension, the number of evaluable wallets during the training phase was 602, 548, 587, 557, 479, and 418, respectively. Methodologically, we first identified 33 overlapping behavioral labels based on predefined economic semantic rules, and then performed K-Means unsupervised clustering on each of the six dimensions. The number of clusters was jointly determined by silhouette coefficients, minimum cluster size constraints, wallet subsampling, and parent event block resampling stability to avoid directly merging behavioral features with different economic meanings into a single global classification. Ultimately, 3, 2, 2, 3, 2, and 2 subgroups were obtained for each of the six dimensions, totaling 14 interpretable behavioral subgroups. The silhouette coefficients for each dimension ranged from 0.327 to 0.501, and the ARI for parent event resampling ranged from 0.631 to 0.814.

As shown in Table \ref{fig:smart_money_behavior}, experimental results show that candidate smart wallets exhibit significant behavioral heterogeneity. Among the corresponding evaluable samples, 40.0\% of wallets showed a preference for high-price entry, 51.3\% exhibited domain-focused behavior, 25.9\% showed independent batch accumulation characteristics, and 43.5\% showed late-stage entry dominance. Cross-week validation further revealed varying degrees of persistence across different behaviors: the retention rate for the relatively high-frequency trading label reached 89.4\%, while the low-price entry label only maintained a retention rate of 34.8\%. These results indicate that candidate smart wallets cannot be considered as a homogeneous trading group, and the six-dimensional profile provides a more granular and interpretable representation of behavior. 

\begin{figure*}[t]
\centering
\begingroup
\setlength{\fboxsep}{5pt}

% ==================================================
% CASE I: POLITICAL PREDICTION MARKET
% ==================================================
\begin{minipage}[t]{0.488\textwidth}
\vspace{0pt}
\noindent\fbox{%
\begin{minipage}[t]{
\dimexpr\linewidth-2\fboxsep-2\fboxrule\relax}
\raggedright\footnotesize
\setlength{\parskip}{2pt}

\textbf{Case I: Politics --- Vance 2028 (GollumGekko)}
\par

\textbf{Wallet:}
\href{https://polygonscan.com/address/0x08458f7e9d2858027de579e4c3ca305475496b6f}
{\texttt{0x08458f7e9d2858027de579e4c3ca305475496b6f}}
\par

\textbf{Observed trade:}
Sep.~29, 2026, 14:41 UTC;
BUY 873.36 Yes @ 0.507 USDC (\$442.79),
increasing holdings from 4,892.24 to 5,765.60 shares.
\href{https://polygonscan.com/tx/0x177fc3cbbf06c229a595c15d312d9dd004a85253f0175a5240ea9f3279463e3f}
{\textbf{Polygon Tx:} \texttt{0x177fc3...63e3f}}.
\par

\textbf{Wallet behavior:}
129 observed political-sector trades
(64 buys, 65 sells) across 27 markets.
The focal buy is 3.46$\times$ the \$127.87 median
of 63 measured buys.
\par

\textbf{Market:}
\href{https://polymarket.com/event/republican-presidential-nominee-2028/will-jd-vance-win-the-2028-republican-presidential-nomination}
{2028 Republican nomination (Vance Yes)};
resolution requires winning and accepting
the nomination.
\par

\textbf{Real-world evidence:}
\href{https://www.pbs.org/newshour/politics/erika-kirk-backs-jd-vances-potential-2028-presidential-bid-at-turning-point-usa-conference}
{PBS/AP}
reports Erika Kirk's personal endorsement;
it does not establish a formal TPUSA
organizational endorsement.
\par

\textbf{MAS trace:}
WalletAnalyst + RulesAnalyst + PoliticsAgent
$\rightarrow$ Skeptic flags personal/organizational
attribution
$\rightarrow$ ClaimVerifier.
\par

\end{minipage}%
}
\end{minipage}
\hfill%
% ==================================================
% CASE II: ESPORTS PREDICTION MARKET
% ==================================================
\begin{minipage}[t]{0.488\textwidth}
\vspace{0pt}
\noindent\fbox{%
\begin{minipage}[t]{
\dimexpr\linewidth-2\fboxsep-2\fboxrule\relax}
\raggedright\footnotesize
\setlength{\parskip}{2pt}

\textbf{Case II: Esports --- Gen.G Worlds 2026 (Anjun)}
\par

\textbf{Wallet:}
\href{https://polygonscan.com/address/0x43372356634781eea88d61bbdd7824cdce958882}
{\texttt{0x43372356634781eea88d61bbdd7824cdce958882}}
\par

\textbf{Observed trade:}
Oct.~11, 2026, 17:46 UTC;
BUY 594.11 Yes @ 0.30 USDC (\$178.23),
increasing holdings from 5,999.41 to
6,593.52 shares.
\emph{Individual Tx hash not present
in supplied case excerpt.}
\par

\textbf{Wallet behavior:}
49 observed LoL trades
(44 buys, 5 sells); median buy \$9.63.
The focal purchase is 18.51$\times$
that sector-sample median.
\par

\textbf{Market:}
\href{https://polymarket.com/event/worlds-2026-winner/will-genptg-win-worlds-2026}
{Gen.G to win Worlds 2026};
settlement follows the championship result.
The \emph{post-trade} Oct.~12 snapshot shows
a Yes bid/ask of 0.30/0.31,
\par

\textbf{Real-world evidence:}
\href{https://www.hotspawn.com/league-of-legends/news/worlds-2026-top-20-players}
{Hotspawn}
commented on Canyon's improving form;
this is third-party analysis,
not an official Riot Games result.
\par

\textbf{MAS trace:}
WalletAnalyst + MarketPanelAnalyst +
RulesAnalyst + EsportsAgent
$\rightarrow$ Skeptic distinguishes
commentary from resolution evidence
$\rightarrow$ ClaimVerifier.
\par

\end{minipage}%
}
\end{minipage}

\endgroup

\caption{Two archived MAS research traces.
Linked wallet, transaction (when available),
market, and source records support inspection.
Internal verification counts are not independent
accuracy estimates.}
\label{fig:mas_evidence_examples}
\end{figure*}

\subsection{Agent Output Example}
\label{sec:agent_output_example}

We examine two archived fills to illustrate how MAS links
wallet-level behavior, a specific on-chain trade, market
semantics, and external evidence. Each case is routed to a
market-specific analyst and reviewed by the Skeptic and
ClaimVerifier. \textbf{Figure~\ref{fig:mas_evidence_examples}} shows
an abbreviated, evidence-grounded decision trace rather
than an unrestricted narrative generated by the model.

\subsection{Trading System and Results}
We developed a smart-money copy-trading strategy for prediction markets. We evaluated the strategy through paper-trading and live-trading experiments on the polymarket.

\begin{table*}[t]
\centering
\caption{Representative smart-money wallets and matched on-chain copy trades.}
\label{tab:smart_money_unified}
\fontsize{6}{6.8}\selectfont
\setlength{\tabcolsep}{0.85pt}
\renewcommand{\arraystretch}{1.13}

\begin{tabular}{@{}l l r r r r l r l l r r r l r r r r r@{}}
\toprule
\multicolumn{6}{c}{\textbf{Smart-money wallet}} &
\multicolumn{7}{c}{\textbf{Original on-chain transaction}} &
\multicolumn{6}{c}{\textbf{Our live follower execution}} \\
\cmidrule(lr){1-6}
\cmidrule(lr){7-13}
\cmidrule(l){14-19}

Address & Specialization & Trades & W & M & A &
Source Tx & Block & Market / Outcome & Act. &
Price & USDC & Shares &
Live Tx & Block & Fill & Limit & Fee & Delay \\
\midrule

\href{https://polygonscan.com/address/0x032eb1bc893940263ad0b01889f262fc232f2a9e}{0x032e..2a9e} &
Sports / Soccer & 19,457 & +7.01 & +35.11 & -2160.63 &
\href{https://polygonscan.com/tx/0x812494e85b8af0f6e4aa1fb68c0b86efe9594fd91120530fdc6e5b0f1ce773cd}{0x812494..} &
95813690 &
\href{https://polymarket.com/event/sea-mon-cag-2026-10-19-more-markets}{Monza--Cagliari O0.5} &
OPEN & 0.9200 & 1142.64 & 1242.00 &
\href{https://polygonscan.com/tx/0xa99db048d7f033020c82687c16fca545a1d223ea7895036260abd5a2f4c3824d}{0xa99db0..} &
95813731 & 0.9200 & 0.94 & 0.0040 & 62 \\

\href{https://polygonscan.com/address/0x076daa87c4fe1a85402a9b6b8e0a866224388d4c}{0x076d..8d4c} &
Sports / Basketball & 47,725 & -26.17 & +552.49 & +3176.54 &
\href{https://polygonscan.com/tx/0x6ff4a5e327b1c79c4b341e8c6f4382f9bbf41f8fea38f68f847ec27f71814c28}{0x6ff4a5..} &
95803028 &
\href{https://polymarket.com/event/bkcba-bei-gua-2026-10-20}{Beijing--Guangdong} &
OPEN & 0.3800 & 11.90 & 31.32 &
\href{https://polygonscan.com/tx/0x8d3119d3d32e7ab178999a9298b41881bd3f00112f13bd9a01356c59407e2728}{0x8d3119..} &
95803153 & 0.3800 & 0.43 & 0.0310 & 188 \\

\href{https://polygonscan.com/address/0x6b97ebe911d7f2e5e3e8c4a0faad4a5c7c0b8d30}{0x6b97..8d30} &
Esports / Dota, CS & 6,329 & +30.46 & +127.23 & +246.51 &
\href{https://polygonscan.com/tx/0x687e3a92a195a2fa83b2c7d689673d4f6562b27074845b39cc4b1bc6b4c9c6ce}{0x687e3a..} &
95803037 &
\href{https://polymarket.com/event/dota2-huliga-nemiga-2026-10-19}{HULIGANI--Nemiga} &
OPEN & 0.7000 & 84.25 & 120.36 &
\href{https://polygonscan.com/tx/0x771bd9ba22f1a3302904aa02227c6aa36ee140c1a51fde73b61fb14052759873}{0x771bd9..} &
95803109 & 0.7100 & 0.75 & 0.0145 & 107 \\

\href{https://polygonscan.com/address/0x83720820a8aa6c3f20ad71850e7a1a17d16c5223}{0x8372..5223} &
Sports / Soccer & 3,396 & +532.25 & +492.74 & +2199.21 &
\href{https://polygonscan.com/tx/0x50c5fd1794d3464c2fe2b205f13ff459041cc2e8706575137cd3431a03ac6c91}{0x50c5fd..} &
95803894 &
\href{https://polymarket.com/event/epl-tot-cov-2026-10-19-more-markets}{Tottenham--Coventry U2.5} &
ADD & 0.4000 & 98.92 & 247.30 &
\href{https://polygonscan.com/tx/0x271388a74b2ae5e79c30695663f5a4b2c619a3008b0e03415df810e72f30db99}{0x271388..} &
95803924 & 0.4100 & 0.45 & 0.0295 & 46 \\

\href{https://polygonscan.com/address/0x85a85805bcbdf2e629788adac1420b6b53949281}{0x85a8..9281} &
Politics / Iran & 154 & +2.29 & +37.31 & -3.88 &
\href{https://polygonscan.com/tx/0x8ddff8414ab4686033acd8db9b433402932eeb524fefde8ccca72c92346f1df8}{0x8ddff8..} &
95814854 &
\href{https://polymarket.com/event/us-iran-ceasefire-continues-throughptptpt}{US--Iran ceasefire Oct31} &
ADD & 0.6600 & 354.35 & 536.90 &
\href{https://polygonscan.com/tx/0xff047f90fdf84ffe13403d319339968c2820574896f64aa085dd0a99c6c89b61}{0xff047f..} &
95814929 & 0.6800 & 0.71 & 0.0000 & 113 \\

\href{https://polygonscan.com/address/0x8d9f8127677a6c64219b1ba5ecd126043551787a}{0x8d9f..787a} &
Esports / LoL & 1,047 & +7.15 & +62.82 & +100.28 &
\href{https://polygonscan.com/tx/0xca66c7c2232c6549f5ad13cc3213415fff565d55c48a8ae603b322014f506450}{0xca66c7..} &
95803061 &
\href{https://polymarket.com/event/lol-g2-dk-2026-10-23}{G2 Esports--Dplus KIA} &
OPEN & 0.4000 & 2000.00 & 5000.00 &
\href{https://polygonscan.com/tx/0x3b39c8a7c7fcf2857723da36157ef420e8d6e66d83d63d844f542be9ad3fc165}{0x3b39c8..} &
95803176 & 0.4000 & 0.45 & 0.0300 & 171 \\

\href{https://polygonscan.com/address/0xa2cced8bfae7d645f7f437fb76becf2fcbb70cbc}{0xa2cc..0cbc} &
Sports / Basketball & 726 & +11.03 & +485.85 & -904.49 &
\href{https://polygonscan.com/tx/0x908acf87db49fad1bb5e3584c04ac9ae1582efcdace73a306a6b0782fcc8d33a}{0x908acf..} &
95815456 &
\href{https://polymarket.com/event/wnba-atl-gsv-2026-10-19}{Atlanta Dream--Valkyries} &
OPEN & 0.4400 & 434.54 & 987.59 &
\href{https://polygonscan.com/tx/0x0f28af78514fd2a12815c065fd92e24393f2d32f303f5a5df9589946b90eef4b}{0x0f28af..} &
95815521 & 0.4500 & 0.49 & 0.0275 & 97 \\

\href{https://polygonscan.com/address/0xcb0a98f88dba12fa101313e060594a34657d4887}{0xcb0a..4887} &
Crypto / BTC intraday & 273 & -12.97 & +0.48 & -1.52 &
\href{https://polygonscan.com/tx/0xdda9b013c7ba5245ab2abfc7a6ee1c32cef581079cd05e0109d5d16984b0507b}{0xdda9b0..} &
95813702 &
\href{https://polymarket.com/event/btc-updown-4h-1792425600}{BTC Oct19, 4h Up} &
OPEN & 0.4965 & 193.87 & 390.50 &
\href{https://polygonscan.com/tx/0xc26923627ac1fa7206f9761ab13556670007bed8118472a84096c6197150b5a1}{0xc26923..} &
95813771 & 0.4783 & 0.54 & 0.0365 & 103 \\

\href{https://polygonscan.com/address/0xd218e474776403a330142299f7796e8ba32eb5c9}{0xd218..b5c9} &
Politics / US post count & 132,846 & -21.96 & +142.71 & +499.05 &
\href{https://polygonscan.com/tx/0x3209ffa588d850d1a3ece8dc24ad0d1b86aed87b8b7385e03fe7143490c67c56}{0x3209ff..} &
95872336 &
\href{https://polymarket.com/event/donald-trump-of-truth-social-posts-october-16-october-23-2026}{Trump posts 160--179 No} &
OPEN & 0.8100 & 16.20 & 20.00 &
\href{https://polygonscan.com/tx/0x751faef0124a1dc423ad406746dc8167808a5118434fa329a0d067e7394fb2e4}{0x751fae..} &
95872412 & 0.8320 & 0.86 & 0.0084 & 114 \\

\href{https://polygonscan.com/address/0xf184bff6a9217f1a76dcba7c0c4888351ebbc2d7}{0xf184..c2d7} &
Esports / LoL & 1,359 & -6.83 & +4.06 & +125.00 &
\href{https://polygonscan.com/tx/0xa096dd12d027d8e3b9f0574f9a5a78610b92976e93e26ad5f59d7f235e7e8276}{0xa096dd..} &
95872817 &
\href{https://polymarket.com/event/lol-los-t1-2026-10-23}{LoL: LOS--T1 / T1} &
ADD & 0.8700 & 59.16 & 68.00 &
\href{https://polygonscan.com/tx/0x5d8c478e9a33001d43562b4642a5ad20c241370e9a31063231a9549160cf4164}{0x5d8c47..} &
95872905 & 0.8800 & 0.92 & 0.0060 & 131 \\

\bottomrule
\end{tabular}

\vspace{3pt}
\parbox{\textwidth}{\fontsize{7}{8.2}\selectfont
\emph{Notes.}
W/M/A denote source-reported sector PnL in USD thousands
(weekly, monthly, and all-time).
Trades are observed wallet activity, not independent orders.
Source USDC is the executed notional
(price multiplied by acquired outcome shares), excluding fees.
Follower trades are confirmed BUY fills;
Fill and Limit are outcome-share prices,
Fee is in USDC, and Delay is in seconds
from source execution to follower on-chain confirmation.
Transaction hashes and markets are clickable links.
}
\end{table*}

\subsubsection{Smart-Money Copy Trading Strategy}
\label{subsec:copy_strategy}

We develop a conditional smart-money copy-trading strategy that monitors on-chain trades and position changes, following only signals that satisfy source profitability, trading-trigger, and directional-position requirements. Instead of replicating the source wallet's trade size, the strategy submits fixed-size Fill-or-Kill (FOK) limit orders of 1 USDC (live) or 1 pUSD (paper)  within  300 seconds  of the source trade. Entry prices are restricted to  [0.05, 0.95) , with a maximum premium of  0.05  above the source price, subject to order-book liquidity, transaction fees, and risk limits. When the source reduces or closes a position, the follower adjusts its holdings proportionally; remaining positions are settled using official outcomes. Independent paper-trading and live-trading executions are recorded to evaluate practical execution feasibility. Details are in the Appendix \ref{trading-system}.

\subsubsection{Paper \& Live Trading Results}
\label{sec:paper_live_results}

As shown in the \textbf{Part C of Figure \ref{fig:system_overview}}, to evaluate the practical feasibility of smart-money
copy trading, we developed a real-time trading framework
supporting both paper trading and live on-chain execution.
The system continuously monitors candidate smart-money
wallets and identifies copy-trading opportunities based on
historical wallet performance, observed trading signals,
and predefined risk constraints. In paper trading, orders
are simulated against the order book available at execution
time, accounting for market depth, limit prices, and
transaction fees. In parallel, the live follower submits
actual on-chain orders with limited capital. Both systems
maintain independent accounts to record executions,
position changes, realized and unrealized PnL, and
settlement or redemption outcomes.

\begin{figure}[t]
    \centering
    \includegraphics[width=0.99\linewidth]
    {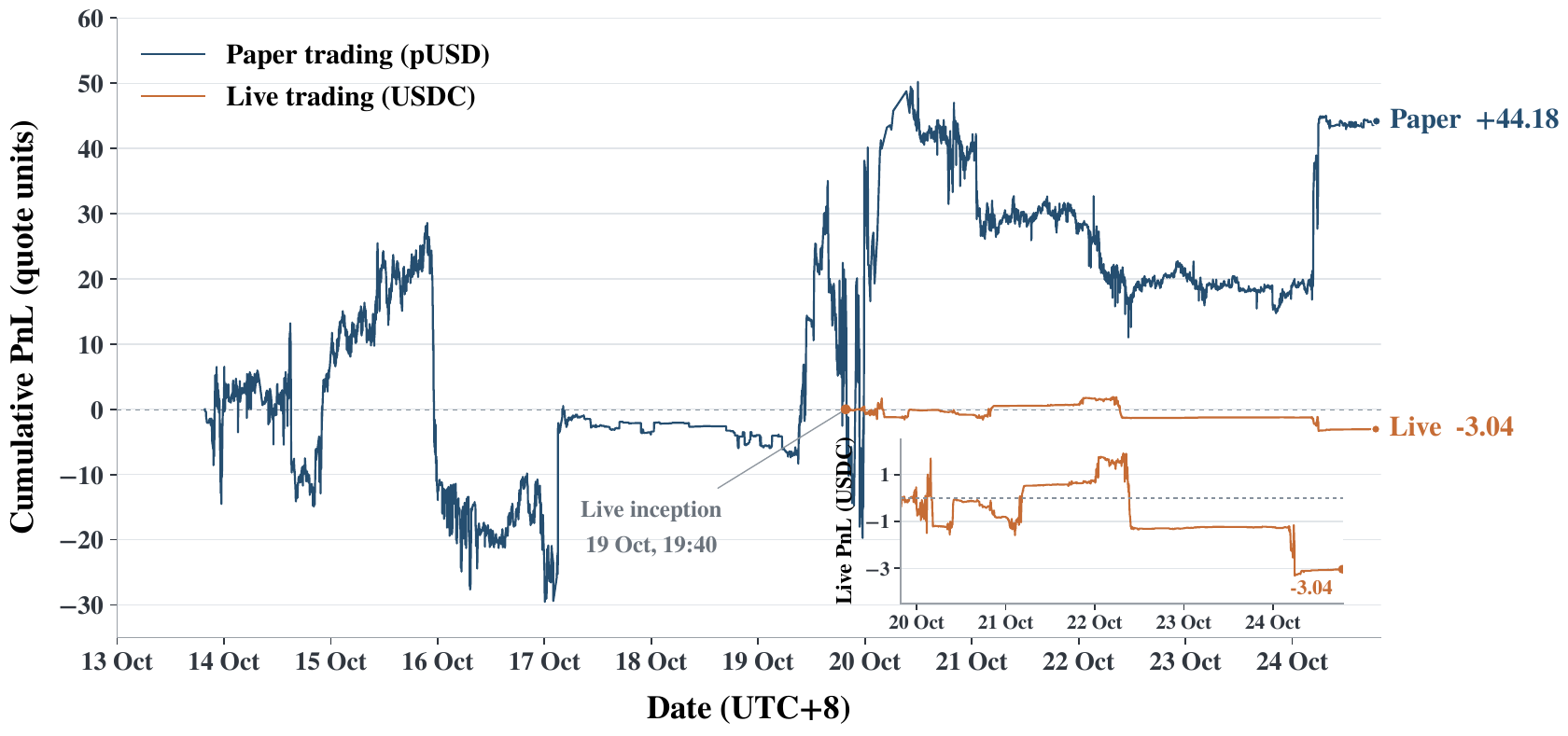}
\caption{Cumulative profit and loss (PnL)
of paper and live copy trading from
October 13 to 24, 2026 (UTC+8).
Paper trading (blue, pUSD) ends at +44.18,
while live trading (orange, USDC), initiated
on October 19, ends at -3.04.
The inset enlarges the live-trading PnL.
The two series differ in observation periods,
trade selection, and accounting units.}
    \label{fig:paper_live_pnl}
\end{figure}

Table~\ref{tab:smart_money_unified} presents ten
representative source wallets and their matched
on-chain follower trades across sports, esports,
politics, and cryptocurrency markets, including
execution prices, fees, and confirmation delays.
Figure~\ref{fig:paper_live_pnl} illustrates the
cumulative PnL of paper and live trading from
October 13 to 24, 2026 (UTC+8).
The main paper account records 283 BUY entries
and a total profit of 44.18 pUSD, whereas
20 live purchases result in a loss of 3.04 USDC.
This divergence is primarily associated with
differences in trade selection and capital allocation.
Specifically, 44 earlier, larger paper trades
contribute +75.73 pUSD, while 239 smaller trades
lose 31.55 pUSD. The single most profitable ETH
trade generates +56.07 pUSD but was not executed
live because of a data-validity check \cite{polymarket2026eth2400,polygonscan2026ethsource}; excluding
this trade reduces aggregate paper PnL to
-11.89 pUSD. Moreover, among 18 commonly closed
trades, matched paper and live executions produce
similar losses of 3.23 pUSD and 3.19 USDC,
respectively. Thus, the observed profitability
gap mainly reflects differences in executed
signals and position sizing, rather than
demonstrating systematic paper--live execution
bias. These results confirm the feasibility
of real-time on-chain copy trading but do not
establish consistent profitability.
Further analysis of execution costs and
profit concentration is presented in
Section~\ref{sec:economic_performance}.

\subsection{Economic Performance}
\label{sec:economic_performance}

\noindent\textbf{Copy-Trading Replication Gap.}
Historical profitability does not guarantee
equivalent follower returns. When a source wallet
purchases outcome shares, it may consume the
best available ask liquidity, forcing subsequent
followers to execute at higher prices.
This execution disadvantage is related to
implementation shortfall and market impact
\cite{perold1988implementation,almgren2001optimal}.
For a nondecreasing ask-price function $a(x)$,
a source purchase of $q_S$ shares followed by
$q_F$ shares satisfies, under a static order book,
\begin{equation}
\bar p_F^{\,0}
=\frac{1}{q_F}\int_{q_S}^{q_S+q_F}a(x)\,dx
\geq
\frac{1}{q_S}\int_0^{q_S}a(x)\,dx
=\bar p_S.
\label{eq:static_copy_cost}
\end{equation}
Unlike the deterministic imitation penalty
in bonding-curve markets \cite{luo2026resisting},
this inequality need not hold in a dynamic CLOB,
where orders are continuously replenished or
repriced \cite{akey2026tradesquotes}.
For one share of the same binary outcome held
to settlement, the net profit difference is
\begin{equation}
\pi_i^S-\pi_i^F
=(p_i^F-p_i^S)+(c_i^F-c_i^S),
\label{eq:copy_profit_gap}
\end{equation}
where $p$ and $c$ represent entry prices and
per-share transaction costs. Thus, a higher
entry price mechanically reduces the follower's
per-share profit, even when both select the
correct outcome.

\noindent\textbf{Execution and Delay.}
Following studies of imitation and delayed
trading information
\cite{oliphant2026return,yang2022financialadvice},
we match source transactions with follower
executions from October 13--22, 2026 (UTC+8),
tracking outcomes through October 24.
Among 20 live purchases, 13 execute above
the source price, five at approximately the
same price, and two below it. The mean and
median price gaps are $+0.00619$ and $+0.01000$
USDC per share, respectively.
Ten representative trades show confirmation
gaps of 46--188 seconds. However, these
intervals do not measure exact CLOB matching
latency, and the observed price differences
cannot be attributed solely to execution delay.

\begin{table}[t]
\centering
\caption{Execution quality and economic performance
of the fixed copy-trading entry cohort.}
\label{tab:copy_economic_results}
\footnotesize
\setlength{\tabcolsep}{2.5pt}
\renewcommand{\arraystretch}{1.08}

\begin{tabular*}{\columnwidth}
{@{\extracolsep{\fill}}lrr@{}}
\toprule
Metric & Paper & Live \\
\midrule
\multicolumn{3}{@{}l}{\emph{Execution quality}} \\
Executed BUYs & 283 & 20 \\
Recorded entry candidates$^{a}$ & 2,896 & 1,384 \\
Submitted FOK BUYs & 713 & 20 \\
Candidate-to-fill rate (\%) & 9.77 & 1.45 \\
Submitted FOK fill rate (\%)$^{b}$ & 39.69 & 100.00 \\
Higher/equal/lower price$^{c}$ & 201/61/21 & 13/5/2 \\
Mean price gap/share & +0.01015 & +0.00619 \\
Median price gap/share & +0.01000 & +0.01000 \\
Median confirmation gap (s)$^{d}$ & -- & 112 \\
\midrule
\multicolumn{3}{@{}l}{\emph{Economic performance}} \\
Entry cost & 693.46 & 20.36 \\
Realized PnL & +39.11 & -3.19 \\
Unrealized PnL & +5.07 & +0.15 \\
Total PnL & +44.18 & -3.04 \\
Paired closed trades ($n=18$)$^{e}$ & -3.23 & -3.19 \\
\midrule
\multicolumn{3}{@{}l}{\emph{Profit concentration}} \\
Larger-entry PnL ($n=44$) & +75.73 & -- \\
Smaller-entry PnL ($n=239$) & -31.55 & -- \\
Largest-event PnL & +56.07 & -- \\
PnL excluding largest event & -11.89 & -- \\
\bottomrule
\end{tabular*}

\vspace{2pt}
\parbox{\columnwidth}{\scriptsize
\emph{Notes.} Monetary values are in pUSD
(paper) or USDC (live); price gaps are per
outcome share. TBD denotes metrics requiring
further computation from archived execution
records. PnL includes recorded fees and
available open-position valuations.
Valuations are not a globally atomic snapshot.
}
\vspace{-0.3cm}
\end{table}

\noindent\textbf{Profitability Implications.}
As shown in Table~\ref{tab:copy_economic_results},
paper trading records $+44.18$ pUSD, but returns
are highly concentrated: larger entries contribute
$+75.73$ pUSD, whereas smaller entries lose
$31.55$ pUSD. The largest event alone contributes
$+56.07$ pUSD; excluding it reverses total
PnL to $-11.89$ pUSD. Live trading loses
$3.04$ USDC, while 18 commonly closed
paper/live pairs both generate losses of
approximately 3.2 in their respective units.
These results show that the copy-trading profitability depends on execution
costs, position sizing, and payoff concentration
rather than source-wallet performance alone.

\section{Conclusion}

In this paper, we conduct a systematic study of smart money
and copy trading in prediction markets using Polymarket.
We reconstruct wallet-level trading activities and examine
historical profit concentration and short-term profitability
persistence, while six-dimensional behavioral clustering
reveals heterogeneous trading patterns. We further develop
an evidence-grounded multi-agent framework that integrates
wallet histories, market conditions, and real-world
information to generate traceable and verifiable trading
explanations. Through order-book-aware paper trading and
live on-chain execution, we evaluate the practical
replicability of smart-money strategies under market
frictions. Overall, our findings demonstrate that
historical profitability does not necessarily translate
into persistent trading advantages or profitable copy
trading, highlighting the importance of behavioral
understanding, evidence verification, and realistic
execution constraints.

%% Camera-ready only. The anonymous option drops this environment.
%% \begin{acks}
%% \end{acks}

%% Add when the work uses human participants or sensitive Web data,
%% including IRB approval numbers where applicable.
%% \section{Ethical Considerations}

\newpage

\bibliographystyle{ACM-Reference-Format}
\bibliography{references}

@misc{pumpfun,
  author = {{Pump.fun}},
  title = {{Pump.fun}: Launch and Trade Memecoins on Solana},
  year = {n.d.},
  url = {https://pump.fun/},
  note = {Accessed: 2026-10-17}
}

@misc{aveai,
  author = {{Ave.ai}},
  title = {{Ave.ai}: On-Chain Trading and Analytics Platform},
  year = {n.d.},
  url = {https://ave.ai/},
  note = {Accessed: 2026-10-17}
}

@misc{xstocks,
  author = {{xStocks}},
  title = {{xStocks}: Tokenized Equities and ETFs},
  year = {n.d.},
  url = {https://xstocks.fi/},
  note = {Accessed: 2026-10-17}
}

@misc{hyperliquid,
  author = {{Hyperliquid}},
  title = {{Hyperliquid}: Infrastructure to House All Finance},
  year = {n.d.},
  url = {https://hyperliquid.xyz/},
  note = {Accessed: 2026-10-17}
}

@misc{polymarket,
  author = {{Polymarket}},
  title = {{Polymarket}: Prediction Markets},
  year = {n.d.},
  url = {https://polymarket.com/},
  note = {Accessed: 2026-10-17}
}

@misc{predictfun,
  author = {{Predict}},
  title = {{Predict}: The BNB-Native Prediction Market},
  year = {n.d.},
  url = {https://predict.fun/},
  note = {Accessed: 2026-10-17}
}

@misc{kalshi,
  author = {{Kalshi}},
  title = {{Kalshi}: Prediction Market for Trading the Future},
  year = {n.d.},
  url = {https://kalshi.com/},
  note = {Accessed: October 17, 2026}
}

@misc{polygon,
  author = {{Polygon Labs}},
  title = {{Polygon Chain Overview}},
  year = {n.d.},
  url = {https://docs.polygon.technology/pos/overview/},
  note = {Accessed: October 17, 2026}
}

@misc{gmgn,
  author = {{GMGN.AI}},
  title = {{GMGN.AI}: Multi-Chain Meme Trading Terminal},
  year = {n.d.},
  url = {https://gmgn.ai/},
  note = {Accessed: October 17, 2026}
}

@misc{hypertracker,
  author = {{HyperTracker}},
  title = {{HyperTracker}: Hyperliquid Wallet and Whale Tracker},
  year = {n.d.},
  url = {https://hypertracker.io/},
  note = {Accessed: October 17, 2026}
}

@misc{hyperliquidanalytics,
  author = {{Hyperliquid Analytics}},
  title = {{Hyperliquid Analytics}: On-Chain Trader Intelligence Terminal},
  year = {n.d.},
  url = {https://hyperliquidanalytics.com/},
  note = {Accessed: October 17, 2026}
}

@misc{polysights,
  author = {{Polysights}},
  title = {{Polysights}: The Intelligence Layer for Prediction Markets},
  year = {n.d.},
  url = {https://polysights.xyz/},
  note = {Accessed: October 17, 2026}
}

@misc{polybeats,
  author = {{BlockBeats}},
  title = {{BlockBeats Pro}: PolyBeats Prediction Market Intelligence},
  year = {n.d.},
  url = {https://en.theblockbeats.news/pro},
  note = {Accessed: October 17, 2026}
}

@inproceedings{luo2026resisting,
  title={Resisting Manipulative Bots in Meme Coin Copy Trading: A Multi-Agent Approach with Chain-of-Thought Reasoning},
  author={Luo, Yichen and Feng, Yebo and Xu, Jiahua and Liu, Yang},
  booktitle={Proceedings of the ACM Web Conference 2026},
  pages={3414--3425},
  year={2026}
}

@article{reichenbach2025exploring,
  title={Exploring decentralized prediction markets: Accuracy, skill, and bias on Polymarket},
  author={Reichenbach, Felix and Walther, Martin},
  journal={Skill, and Bias on Polymarket (December 12, 2025)},
  year={2025}
}

@misc{umaoracle,
  author = {{UMA Protocol}},
  title = {{How Does UMA Work?}},
  year = {n.d.},
  url = {https://docs.uma.xyz/protocol-overview/how-does-umas-oracle-work},
  note = {Accessed: October 18, 2026}
}

@article{szwajcok2026meme,
  title={Meme Coin Factories: Uncovering Large-Scale Manipulations on pump. fun},
  author={Szwajcok, Nicolas and Tsuchiya, Taro and Liu, Enze and Soska, Kyle and Payer, Mathias and Christin, Nicolas},
  journal={arXiv preprint arXiv:2609.10246},
  year={2026}
}

@article{oliphant2026return,
  title={The Return to Imitation: Evidence from Copy Trading on Polymarket},
  author={Oliphant, Patrick},
  journal={Available at SSRN 6670318},
  year={2026}
}

@inproceedings{li2024cryptotrade,
  title={CryptoTrade: A reflective LLM-based agent to guide zero-shot cryptocurrency trading},
  author={Li, Yuan and Luo, Bingqiao and Wang, Qian and Chen, Nuo and Liu, Xu and He, Bingsheng},
  booktitle={Proceedings of the 2024 Conference on Empirical Methods in Natural Language Processing},
  pages={1094--1106},
  year={2024}
}

@article{luo2025llm,
  title={Llm-powered multi-agent system for automated crypto portfolio management},
  author={Luo, Yichen and Feng, Yebo and Xu, Jiahua and Tasca, Paolo and Liu, Yang},
  journal={arXiv preprint arXiv:2501.00826},
  year={2025}
}

@article{wei2025advanced,
  title={Advanced smart contract vulnerability detection via llm-powered multi-agent systems},
  author={Wei, Zhiyuan and Sun, Jing and Sun, Yuqiang and Liu, Ye and Wu, Daoyuan and Zhang, Zijian and Zhang, Xianhao and Li, Meng and Liu, Yang and Li, Chunmiao and others},
  journal={IEEE Transactions on Software Engineering},
  year={2025},
  publisher={IEEE}
}

@inproceedings{wang2026cogagent,
  title={CogAgent: Self-Evolving Cognitive Agents for Multi-Source Fraud Detection in Heterogeneous Financial Networks},
  author={Wang, Qianyu and Tian, Weiyou and Wang, Rong and Tsai, Wei-Tek and Shi, Tianyu and Liu, Zhuang and Xia, Tianze},
  booktitle={Proceedings of the ACM Web Conference 2026},
  pages={3449--3458},
  year={2026}
}

@inproceedings{zhong2025detecting,
  title={Detecting Various DeFi Price Manipulations with LLM Reasoning},
  author={Zhong, Juantao and Wu, Daoyuan and Liu, Ye and Xie, Maoyi and Liu, Yang and Li, Yi and Liu, Ning},
  booktitle={2025 40th IEEE/ACM International Conference on Automated Software Engineering (ASE)},
  pages={1781--1793},
  year={2025},
  organization={IEEE}
}

@inproceedings{muzammil2025poorest,
  title={The poorest man in babylon: A longitudinal study of cryptocurrency investment scams},
  author={Muzammil, Muhammad and Pitumpe, Abisheka and Li, Xigao and Rahmati, Amir and Nikiforakis, Nick},
  booktitle={Proceedings of the ACM on Web Conference 2025},
  pages={1034--1045},
  year={2025}
}

@article{niu2025natlm,
  title={Natlm: Detecting defects in nft smart contracts leveraging llm},
  author={Niu, Yuanzheng and Li, Xiaoqi and Li, Wenkai},
  journal={arXiv preprint arXiv:2508.01351},
  year={2025}
}

@misc{polymarketapi,
  author = {{Polymarket}},
  title = {{Polymarket Data API v2 Documentation}},
  year = {n.d.},
  url = {https://docs.polymarket.com/api-reference/data-api/overview},
  note = {Accessed: October 19, 2026}
}

@misc{polygonbor,
  author = {{Polygon Labs}},
  title = {{Bor}: Polygon PoS Execution Client},
  year = {n.d.},
  url = {https://github.com/0xPolygon/bor},
  note = {Official source code repository.
          Accessed: October 19, 2026}
}

@techreport{erc1155,
  author = {Radomski, Witek and Cooke, Andrew
            and Castonguay, Philippe and Therien, James
            and Binet, Eric and Sandford, Ronan},
  title = {{ERC-1155}: Multi Token Standard},
  institution = {Ethereum Improvement Proposals},
  type = {ERC},
  number = {1155},
  year = {2018},
  month = jun,
  url = {https://eips.ethereum.org/EIPS/eip-1155}
}

@misc{qwen38,
    title = {{Qwen3.8-Max}: A New Bar for Coding and Cowork},
    url = {https://qwen.ai/blog?id=qwen3.8},
    author = {{Qwen Team}},
    month = {August},
    year = {2026}
}

@article{gruber1996another,
  author  = {Gruber, Martin J.},
  title   = {Another Puzzle: The Growth in Actively
             Managed Mutual Funds},
  journal = {The Journal of Finance},
  year    = {1996},
  volume  = {51},
  number  = {3},
  pages   = {783--810},
  doi     = {10.1111/j.1540-6261.1996.tb02707.x}
}

@article{zheng1999money,
  author  = {Zheng, Lu},
  title   = {Is Money Smart? A Study of Mutual Fund
             Investors' Fund Selection Ability},
  journal = {The Journal of Finance},
  year    = {1999},
  volume  = {54},
  number  = {3},
  pages   = {901--933},
  doi     = {10.1111/0022-1082.00131}
}

@article{sapp2004momentum,
  author  = {Sapp, Travis and Tiwari, Ashish},
  title   = {Does Stock Return Momentum Explain
             the Smart Money Effect?},
  journal = {The Journal of Finance},
  year    = {2004},
  volume  = {59},
  number  = {6},
  pages   = {2605--2622},
  doi     = {10.1111/j.1540-6261.2004.00710.x}
}

@techreport{akey2026polymarket,
  author      = {Akey, Pat and Grégoire, Vincent
                 and Harvie, Nicolas
                 and Martineau, Charles},
  title       = {Who Wins and Who Loses in
                 Prediction Markets?
                 Evidence from Polymarket},
  institution = {Centre for Economic Policy Research},
  type        = {CEPR Discussion Paper},
  number      = {21615},
  year        = {2026},
  url         = {https://cepr.org/publications/dp21615}
}

@article{fama2010luck,
  author  = {Fama, Eugene F. and French, Kenneth R.},
  title   = {Luck versus Skill in the Cross-Section
             of Mutual Fund Returns},
  journal = {The Journal of Finance},
  year    = {2010},
  volume  = {65},
  number  = {5},
  pages   = {1915--1947},
  doi     = {10.1111/j.1540-6261.2010.01598.x}
}

@article{barras2010false,
  author  = {Barras, Laurent and Scaillet, Olivier
             and Wermers, Russ},
  title   = {False Discoveries in Mutual Fund
             Performance: Measuring Luck in
             Estimated Alphas},
  journal = {The Journal of Finance},
  year    = {2010},
  volume  = {65},
  number  = {1},
  pages   = {179--216},
  doi     = {10.1111/j.1540-6261.2009.01527.x}
}

@article{hong2013poisson,
  author  = {Hong, Yili},
  title   = {On Computing the Distribution Function
             for the Poisson Binomial Distribution},
  journal = {Computational Statistics
             \& Data Analysis},
  year    = {2013},
  volume  = {59},
  pages   = {41--51},
  doi     = {10.1016/j.csda.2012.10.006}
}

@article{benjamini1995controlling,
  author  = {Benjamini, Yoav and Hochberg, Yosef},
  title   = {Controlling the False Discovery Rate:
             A Practical and Powerful Approach
             to Multiple Testing},
  journal = {Journal of the Royal Statistical
             Society: Series B},
  year    = {1995},
  volume  = {57},
  number  = {1},
  pages   = {289--300},
  doi     = {10.1111/j.2517-6161.1995.tb02031.x}
}

@article{benjamini2001control,
  author  = {Benjamini, Yoav and Yekutieli, Daniel},
  title   = {The Control of the False Discovery Rate
             in Multiple Testing under Dependency},
  journal = {The Annals of Statistics},
  year    = {2001},
  volume  = {29},
  number  = {4},
  pages   = {1165--1188},
  doi     = {10.1214/aos/1013699998}
}

@article{perold1988implementation,
  author = {Perold, Andre F.},
  title = {The Implementation Shortfall:
           Paper versus Reality},
  journal = {The Journal of Portfolio Management},
  volume = {14},
  number = {3},
  pages = {4--9},
  year = {1988},
  doi = {10.3905/jpm.1988.409150}
}

@article{almgren2001optimal,
  author = {Almgren, Robert and Chriss, Neil},
  title = {Optimal Execution of Portfolio Transactions},
  journal = {The Journal of Risk},
  volume = {3},
  number = {2},
  pages = {5--39},
  year = {2001},
  doi = {10.21314/JOR.2001.041}
}

@misc{akey2026tradesquotes,
  author = {Akey, Pat and Gr{\'e}goire, Vincent
            and Harvie, Nicolas and Martineau, Charles},
  title = {Trades, Quotes, and Price Discovery
           in Prediction Markets},
  year = {2026},
  doi = {10.2139/ssrn.7547106}
}

@article{yang2022financialadvice,
  author = {Yang, Mingwen and Zheng, Zhiqiang
            and Mookerjee, Vijay},
  title = {How Much Is Financial Advice Worth?
           The Transparency-Revenue Tension
           in Social Trading},
  journal = {Management Science},
  volume = {68},
  number = {7},
  pages = {5252--5268},
  year = {2022},
  doi = {10.1287/mnsc.2021.4147}
}

@misc{polymarket2026eth2400,
  author = {{Polymarket}},
  title = {Will the Price of Ethereum Be Above
           \$2,400 on October 19, 2026?},
  year = {2026},
  note = {Polymarket prediction market,
          Gamma Market ID: 5610574.
          Accessed: October 24, 2026},
  url = {https://polymarket.com/event/ethereum-above-on-october-19-2026/ethereum-above-2400-on-october-19-2026}
}

@misc{polygonscan2026ethsource,
  author = {{PolygonScan}},
  title = {Polygon Transaction
           0x1b97df5943af75bb06840b3215376f8f13321a07172f2fbca415884e44e12b96},
  year = {2026},
  note = {Source-wallet transaction for the
          Polymarket Ethereum price market,
          October 19, 2026.
          Accessed: October 24, 2026},
  url = {https://polygonscan.com/tx/0x1b97df5943af75bb06840b3215376f8f13321a07172f2fbca415884e44e12b96}
}

%% An appendix counts toward the page limit. Reviewers need not
%% read past page 8 of a long paper.
%% \appendix
%% \section{Additional Material}

\newpage
\section{Appendix}
\label{background}
\subsection{Background}

\noindent\textbf{Prediction Market.}
Prediction Markets aggregate dispersed information by trading contracts associated with uncertain future events. In binary prediction markets, Yes and No shares correspond to mutually exclusive event outcomes, with theoretical prices ranging from $0 to $1, and satisfying a price complementarity relationship under frictionless, no-arbitrage conditions. Polymarket is a decentralized prediction market based on Polygon, using an off-chain centralized limit order book (CLOB) for order matching and smart contracts for on-chain transaction settlement. Event outcomes are determined according to predefined rules, typically through a UMA optimistic oracle. The above pricing, trading, and settlement mechanism can be formally represented as:

\begin{equation}
\label{eq:prediction_market_mechanism}
\begin{aligned}
p_Y^*(t)+p_N^*(t)&=1,\\
p_Y^*(t),p_N^*(t)&\in[0,1],\\
m_o(t)&=\operatorname{Match}_{\mathrm{CLOB}}
(\mathcal{B}_o(t),\mathcal{A}_o(t)),\\
m_o(t)&\xrightarrow{\mathrm{Polygon}}
\operatorname{Settle}(m_o(t)),\\
R_o(\omega)&=\mathbb{I}[\omega=o],\\
R_Y(\omega)+R_N(\omega)&=1.
\end{aligned}
\end{equation}

Where $p_Y^*(t)$ and $p_N^*(t)$ represent the theoretical no-arbitrage price, and $m_o(t)$ represents the transaction formed through CLOB buy and sell orders. After the event is adjudicated, the winning share pays \$1, and the losing share is worth zero.

\noindent\textbf{Smart Money and Copy Trading.}
In the cryptocurrency market, "smart money" typically refers to wallets that attract attention due to their historical profitability, trading win rate, or other outstanding performance. On-chain analytics platforms such as GMGN and HyperTracker provide investors with wallet screening, transaction monitoring, and copy trading services by aggregating publicly available transaction records, position changes, and historical returns. However, historical profitability does not necessarily equate to consistent trading ability; returns may stem from expertise, informational advantages, arbitrage opportunities, or random luck. Copy trading refers to investors observing the trading behavior of a target wallet and attempting to replicate its buying and selling directions or position changes to achieve similar investment returns. Because of the delay between signal recognition and trade execution, and because market liquidity, order book depth, spreads, and transaction costs are constantly changing, followers may achieve different net returns even when copying the same trading direction.

\subsection{Experimental Parameters}
\label{experimental-parameters}
Details are shown in the Table \ref{tab:experimental_environment}.

\subsection{Trading System}
\label{trading-system}

\noindent\textbf{Trade Monitoring and Wallet Selection.}
Our system monitors candidate smart-money wallets
using Polymarket APIs and a self-hosted Polygon node.
Confirmed OrderFilled events, ERC-1155 transfers,
and transaction receipts are used to reconstruct
source trades and position changes. The monitor
polls every 2 seconds, processing up to 32 blocks
per batch. Wallet eligibility requires positive
recent domain-specific PnL, known historical
performance, and valid evidence. Experimental
follower eligibility does not require passing
the full 180/90-day smart-money criteria.

\noindent\textbf{Signal Generation.}
We construct follower units at the
wallet--topic--trading-style level.
Consecutive purchases of the same outcome
are grouped within 30 seconds, requiring
at least 10 USDC in source notional.
BUY signals are triggered by alternative
conditions, including (i) purchases exceeding
twice the historical median size,
(ii) position increases of at least 50\%,
(iii) qualifying historical observations,
or (iv) authorized experimental openings.
Historical comparisons use at least 10
purchase groups over the preceding 30 days.
Duplicate signals are filtered before execution.

\noindent\textbf{Order Execution and Risk Control.}
Each follower BUY uses a fixed nominal amount
of 1 pUSD (paper) or 1 USDC (live).
For source price $p_s$, maximum premium
$\delta=0.05$, and tick size $\tau$,
the preliminary price limit is
\begin{equation}
p_{\mathrm{cap}}=
\tau\left\lfloor
\frac{\min(p_s+\delta,\,0.95)}{\tau}
\right\rfloor.
\end{equation}
The final entry price must lie in $[0.05,0.95)$.
Signals expire 300 seconds after the source trade.
Both paper and live execution use Fill-or-Kill
(FOK) orders, subject to market availability,
order-book depth, fees, and account risk limits.
Paper orders use the actual execution-time
order book rather than the source execution price.
Units lacking sufficient historical financial
statistics may enter through a separately
marked experimental path.

\noindent\textbf{Position Exit and Settlement.}
Follower exits replicate the source wallet's
proportional reductions:
\begin{equation}
Q_f^{\mathrm{sell}}
=
Q_f\left(
1-\frac{Q_s^{\mathrm{after}}}
{Q_s^{\mathrm{before}}}
\right),
\end{equation}
where $Q_f$ is the follower's current position.
Exit orders undergo independent liquidity and
position checks. Remaining positions are settled
using verified market outcomes and redemption
records. Paper and live accounts independently
maintain orders, fills, fees, positions, and
realized/unrealized PnL. Idempotent order
identifiers prevent duplicate execution.
MAS research provides trading explanations,
while actual order authorization remains
rule-based and independent of offline
clustering and logistic regression.

% Insert after existing Experimental Parameters / Trading System sections.
% Add \\input{www_mas_concrete_prompts_style.tex} in the preamble.
% Every INPUT here is a condensed presentation of archived records.
% Every OUTPUT here is an archived report excerpt or recorded audit outcome,
% NOT a verbatim raw intermediate SpecialistReport JSON response.
% All prompts below are verbatim SELECTED LINES from the implemented role prompts.

\subsection{Prompts for the MAS}
\label{prompts-for-the-mas}

Each MAS call combines a role-specific system prompt, a shared evidence-constrained user prompt, and a structured response schema. All roles are restricted to the supplied evidence packet and must return schema-compliant JSON objects. Core fields include the claim statement, modality, evidence references, temporal scope, dependencies, and required qualifications. Below we show representative prompt excerpts together with condensed examples from archived runs.

\begin{figure*}[t]
\centering
\scriptsize
\setlength{\tabcolsep}{4pt}
\renewcommand{\arraystretch}{1.02}

\begin{minipage}[t]{0.485\textwidth}
\vspace{0pt}
\begin{smcase}{Prompt 1: Wallet Analysis Agent}
\textbf{Selected instructions.}

``You are WalletAnalysisAgent.''

``Analyze the wallet portrait and recent PnL as separate findings.''

``Return claims about sourced transaction frequency, buy/sell pattern, position changes, and trade-size distribution within the exact sampled sector and interval.''

``ACCOUNT scope is the whole account, not a selected board.''

``Partial samples are allowed if explicitly identified as samples.''

``Do not compute missing win rate, profit factor, or holding time.''

``Mark facts as FACT and explanations as INFERENCE.''

``Do not search, draft publication text, or approve publication.''
\end{smcase}
\end{minipage}
\hfill
\begin{minipage}[t]{0.485\textwidth}
\vspace{0pt}
\begin{smcase}{Example 1: Wallet Research --- Vance 2028}
\textbf{Input.} Wallet \texttt{0x08458f7e...75496b6f} (GollumGekko); BUY 873.36 Yes shares at 0.507 USDC (442.79 USDC), position action: ADD; source tx \texttt{0x177fc3...63e3f}. Observed political-sector sample: 129 trades across 27 markets; median BUY size 127.87 USDC.

\textbf{Condensed output.}
The focal trade is approximately \textbf{3.46$\times$} the median BUY size in the observed political-sector sample. This is a trade-size comparison rather than evidence of private motive or persistent forecasting skill; the sample is not represented as the wallet's complete history.
\end{smcase}
\end{minipage}

\vspace{3pt}

\begin{minipage}[t]{0.485\textwidth}
\vspace{0pt}
\begin{smcase}{Prompt 2: Rules and Politics Analysts}
\textbf{Rules Analyst.}

``Read only the supplied rule text.''

``Identify conditions, objects, dates, time zones, required sources, exceptions, and ambiguities.''

``Do not invent fields, decide settlement, or use the title in place of missing rules.''

\smallskip
\textbf{Politics Agent.}

``Resolve the exact people, office, institution, jurisdiction, and decision stage before reasoning.''

``Rank official records above media interpretation.''

``Form the political assessment before interpreting the wallet trade.''
\end{smcase}
\end{minipage}
\hfill
\begin{minipage}[t]{0.485\textwidth}
\vspace{0pt}
\begin{smcase}{Example 2: Political Evidence --- Vance 2028}
\textbf{Input.} Market: Will J.D.~Vance win the 2028 Republican presidential nomination? The Yes outcome requires winning and accepting the nomination. Archived reporting describes Erika Kirk's endorsement and related political developments.

\textbf{Condensed output.}
Erika Kirk's endorsement is an individual political statement, not verified organizational endorsement by Turning Point USA. Such reporting provides relevant background but does not determine market settlement or establish the wallet's trade-time information set. Later evidence is marked \texttt{RESEARCH\_UPDATE}.
\end{smcase}
\end{minipage}

\vspace{3pt}

\begin{minipage}[t]{0.485\textwidth}
\vspace{0pt}
\begin{smcase}{Prompt 3: Market Panel and Esports Agents}
\textbf{Market Panel Agent.}

``Return sourced claims for odds, bid/ask spread, visible depth, and the actually observed price-history interval.''

``Include market identity, units, and snapshot time.''

``Do not apply today's book to historical execution.''

\smallskip
\textbf{Esports Agent.}

``Analyze the exact title, tournament, format, roster, schedule, and match state.''

``Distinguish scheduled, live, postponed, and final series according to the market rules.''

``Preserve the metric, value, denominator, and reporting period of each statistic.''
\end{smcase}
\end{minipage}
\hfill
\begin{minipage}[t]{0.485\textwidth}
\vspace{0pt}
\begin{smcase}{Example 3: Esports Market --- Gen.G Worlds 2026}
\textbf{Input.} Wallet \texttt{0x43372356...ce958882} (Anjun); BUY 594.11 Yes shares at 0.30 USDC (178.23 USDC), position action: ADD. Observed LoL sample: 49 trades (44 BUY, 5 SELL); purchase size \textbf{18.51$\times$} the sample median. Oct.~12 market snapshot: Yes bid/ask 0.30/0.31.

\textbf{Condensed output.}
The trade added to an existing Gen.G position. The Oct.~12 quote cannot establish the executable book at the Oct.~11 entry time. Third-party commentary neither verifies the eventual result nor reveals the wallet's trading motive. Later-acquired evidence is marked \texttt{RESEARCH\_UPDATE}.
\end{smcase}
\end{minipage}

\vspace{3pt}

\begin{minipage}[t]{0.485\textwidth}
\vspace{0pt}
\begin{smcase}{Prompt 4: Skeptic, Verifier, and Editor}
\textbf{Skeptic.}

``Stress-test proposed signals without deleting supported observations.''

``Each counter-hypothesis must identify the challenged claim, evidence, and required qualification.''

\smallskip
\textbf{Evidence Verifier.}

``Review each claim against raw evidence, not another agent's summary.''

``Mark SUPPORTED only when the cited evidence directly entails the statement at the claimed precision and time scope.''

\smallskip
\textbf{Editor / Draft Verifier.}

``Use only verified claims.''

``Do not add facts, numbers, motives, or causal claims.''

``Recheck the final draft against the same original evidence.''
\end{smcase}
\end{minipage}
\hfill
\begin{minipage}[t]{0.485\textwidth}
\vspace{0pt}
\begin{smcase}{Example 4: Verification Outcomes}
\textbf{Positive case: Vance 2028.}
The Skeptic challenged the inference that Erika Kirk's personal endorsement constituted formal TPUSA support. After source-bound revision, the archived run recorded \textbf{15/15 VERIFIED} claims and \textbf{10/10 SUPPORTED} draft blocks; final validation \textbf{PASS}; publication policy \texttt{REVIEW\_REQUIRED}.

\smallskip
\textbf{Negative case: Fed Rate Decision.}
The archived run recorded \textbf{15/18 VERIFIED} claims and \textbf{3/18 UNSUPPORTED} claims; final research validation \textbf{FAIL}; disposition \texttt{VALIDATION\_FAILED}. Supported individual claims did not override unsupported report-level evidence.
\end{smcase}
\end{minipage}

\caption{Representative prompt excerpts and archived examples from the evidence-grounded multi-agent system. All roles operate under evidence-constrained inputs and schema-constrained outputs. Examples are concise renderings of archived runs rather than raw intermediate model responses.}
\label{fig:mas_prompt_examples_compact}
\end{figure*}

\begin{table*}[t]
\centering
\caption{Hardware, Software Environment, and Experimental
Configurations.}
\label{tab:experimental_environment}

\begingroup
\fontsize{8}{9.6}\selectfont
\setlength{\tabcolsep}{4pt}
\renewcommand{\arraystretch}{1.10}

\begin{tabularx}{\textwidth}{@{}
>{\raggedright\arraybackslash}p{0.15\textwidth}
>{\raggedright\arraybackslash}p{0.22\textwidth}
>{\raggedright\arraybackslash}X@{}}

\toprule
\textbf{Component} &
\textbf{Parameter} &
\textbf{Configuration} \\
\midrule

\multicolumn{3}{@{}l}{\textbf{A. Server Hardware}} \\
\addlinespace[2pt]

Hardware & Operating system &
Ubuntu 22.04.5 LTS (x86\_64);
Linux kernel 6.8.0-124-generic. \\

& CPU &
2 $\times$ AMD EPYC 9554;
128 physical cores and 256 logical threads. \\

& Memory and storage &
1,133.52 GiB RAM ($\approx$1.11 TiB);
NVMe SSD, approximately 1.8 TiB system
and 7.0 TiB data filesystems. \\

& GPU &
8 $\times$ NVIDIA GeForce RTX 5090;
32,607 MiB VRAM per GPU;
driver 580.76.05. \\

& Experimental allocation &
MAS inference: GPUs 4 and 5 (TP=2);
logistic regression and K-Means: CPU. \\

\midrule
\multicolumn{3}{@{}l}{\textbf{B. Software Environment}} \\
\addlinespace[2pt]

Software & Research environment &
Python 3.12.14; NumPy 2.5.4;
pandas 2.3.3; SciPy 1.18.1. \\

& Data analysis &
DuckDB 1.5.5; scikit-learn 1.9.1;
Matplotlib 3.11.2. \\

& MAS client &
Pydantic 2.13.5; requests 2.34.2;
custom Python MasEngine orchestrator. \\

& Model-serving environment &
Python 3.10.19; SGLang 0.5.13.post1;
Transformers 5.8.1. \\

& GPU computation &
PyTorch 2.11.0+cu130;
CUDA 13.0; Triton 3.5.1. \\

\midrule
\multicolumn{3}{@{}l}{\textbf{C. Logistic Regression}} \\
\addlinespace[2pt]

Logistic
regression &
Dataset and validation &
1,101 candidate wallets;
1,088 wallets with 3,207 weekly pairs;
expanding-window temporal validation,
2,156 held-out predictions. \\

& Models &
M0: training base rate;
M1: previous-week profitability;
M2: profitability, trading activity,
and absolute PnL magnitude. \\

& Feature processing &
Log-transformed activity and PnL magnitude;
standardization using training-set statistics;
no regularization. \\

& Optimization &
Maximum likelihood estimation;
SciPy BFGS; zero initialization;
500 maximum iterations;
gradient tolerance $10^{-6}$. \\

& Statistical inference &
Wallet-clustered sandwich standard errors;
95\% Wald confidence intervals;
800 wallet-block bootstrap replications;
seed 20261013. \\

& Evaluation &
Brier score, log loss, AUC,
calibration error (5 bins);
numerical convergence checks. \\

\midrule
\multicolumn{3}{@{}l}{\textbf{D. Six-Dimensional Behavioral Clustering}} \\
\addlinespace[2pt]

K-Means &
Data and windows &
1,101 candidate wallets;
training: Oct. 6--13, 2026;
validation: Oct. 13--20, 2026 (UTC). \\

& Dimensions &
Activity (4), Scale (4), Price (5),
Focus (6), Position (4), Timing (3)
features; evaluable training wallets:
602/548/587/557/479/418. \\

& Feature preprocessing &
A/B/D/E: 1st--99th percentile winsorization,
selected log1p transforms, training-only
standardization;
C/F: square-root transformation
of compositional proportions. \\

& K-Means optimization &
k-means++ initialization;
$n_{\mathrm{init}}=50$;
Lloyd algorithm;
$\mathrm{max\_iter}=300$;
$\mathrm{tol}=10^{-4}$;
seed 20261013. \\

& Cluster selection &
Candidate $K=1,\ldots,5$;
minimum cluster size 10 and 5\%;
silhouette $\geq0.15$;
prefer smaller $K$ within 0.02
of the best eligible silhouette. \\

& Stability verification &
50 wallet subsampling runs (80\%);
50 parent-event bootstrap runs;
median ARI $\geq0.60$;
at least 40 valid runs per test;
frozen cross-week assignment. \\

& Final profiles &
Selected $K=(3,2,2,3,2,2)$;
14 dimension-specific clusters and
33 potentially overlapping behavioral labels. \\

\midrule
\multicolumn{3}{@{}l}{\textbf{E. Multi-Agent System (MAS)}} \\
\addlinespace[2pt]

MAS &
Base model and serving &
Locally served Qwen3.8-27B model alias;
shared weights across agents;
SGLang with tensor parallelism TP=2. \\

& Generation parameters &
Temperature 0.2; thinking disabled;
32,768-token context;
512-token safety margin;
no explicitly recorded generation seed. \\

& Inference server &
Static memory fraction 0.92;
max running requests 1;
chunked prefill size 512;
Triton attention backend;
PyTorch sampling backend. \\

& Output limits &
Typically 2,048 tokens for core agents;
4,096 for domain experts and draft review;
4,096--6,144 for editing;
up to 8,192 for selected verification tasks. \\

& Shared research budget &
100 total requests; 1,800 s per run;
180 s per-request timeout;
100,000 input characters;
2,000,000 response bytes
(Oct. 22 configuration). \\

& Output and verification &
JSON Schema-constrained responses;
Pydantic validation;
evidence-grounded claims;
bounded repair and re-verification. \\

\bottomrule
\end{tabularx}

\vspace{3pt}

\parbox{\textwidth}{
\fontsize{7.5}{9}\selectfont
\emph{Notes.}
Hardware and currently installed software were
inspected on October 24, 2026.
The clustering software versions are independently
recorded in its archived environment manifest;
the earlier logistic-regression experiment lacks
an equally complete environment snapshot.
Eight GPUs are installed, but only two were
allocated to MAS inference.
CPU core counts do not represent actual
per-experiment thread usage.
For numerical experiments,
\texttt{OMP\_NUM\_THREADS=1} and
\texttt{OPENBLAS\_NUM\_THREADS=1}
were used in the clustering configuration.
The MAS model alias does not establish
an exact public checkpoint revision.
Request budgets are run-specific limits,
not measured consumption.
}

\endgroup
\end{table*}

\end{document}